\documentclass[acmsmall]{acmart}
\renewcommand\footnotetextcopyrightpermission[1]{} 

\usepackage{graphicx}
\usepackage{textcomp}
\usepackage{amsmath,amsfonts,amssymb}
\usepackage{algorithm,algorithmic}
\usepackage{graphicx}
\usepackage[pscoord]{eso-pic}
\usepackage{lipsum}
\usepackage{booktabs} 
\usepackage{enumitem}
\usepackage{colortbl} 
\usepackage{multirow}
\usepackage{epstopdf}
\usepackage{diagbox}
\usepackage{svg}
\usepackage{xcolor}
\usepackage{verbatim} 
\usepackage{multicol}
\usepackage{hyphenat}
\usepackage{lipsum}
\usepackage{enumitem}
\usepackage{etoolbox}
\usepackage{mathtools}
\usepackage{euscript}
\usepackage{mathrsfs}
\usepackage{bm}
\usepackage[english]{babel}
\usepackage{blindtext}
\usepackage{textcase}
\usepackage{color}
\usepackage{tikz}
\usepackage{pifont}
\usepackage{makecell}
\usepackage[T1]{fontenc}
\usepackage{hyperref}
\usepackage[acronym]{glossaries}
\usepackage{subcaption}
\usepackage[tablename=TABLE, tableposition=top]{caption}
\usepackage{pifont}
\usepackage{makecell}
\usepackage[T1]{fontenc}
\usepackage{hyperref}
\usepackage[acronym]{glossaries}
\usepackage[nodayofweek]{datetime}
\usepackage[compact]{titlesec}

\usepackage[all]{background}
\SetBgContents{© {ACM} {2023}. This is the author's version of the work. It is posted here for your personal use. Not for redistribution.}
\SetBgScale{0.95}
\SetBgColor{black}
\SetBgAngle{0}
\SetBgOpacity{0.7}
\SetBgPosition{current page.north east}
\SetBgVshift{-0.75cm}
\SetBgHshift{-9.25cm}

\newcommand\freefootnote[1]{%
	\let\thefootnote\relax%
	\footnotetext{#1}%
	\let\thefootnote\svthefootnote%
}

\newcommand{\cmark}{\ding{51}}%
\newcommand{\xmark}{\ding{55}}%

\DeclareMathAlphabet{\mathpzc}{OT1}{pzc}{m}{it}

\definecolor{add}{rgb}{1, 0, 0}
\definecolor{rev}{rgb}{0, 0, 1}
\definecolor{reg}{rgb}{0, 0, 0}

\addto\extrasenglish{%

}
\makeglossaries
\newacronym{axc}{AxC}{\text{A}pproximate \text{C}omputing}
\newacronym{axo}{AxO}{\text{A}pproximate \text{O}perator}
\newacronym{ai}{AI}{\text{A}rtificial \text{I}ntelligence}
\newacronym{cpd}{CPD}{\text{C}ritical \text{P}ath \text{D}elay}
\newacronym{pdp}{PDP}{\text{P}ower-\text{D}elay \text{P}roduct}
\newacronym{ilp}{ILP}{\text{I}nstruction \text{L}evel \text{P}arallelism}
\newacronym{vlsi}{VLSI}{\text{V}ery \text{L}arge \text{S}cale \text{I}ntegration}
\newacronym{mttf}{MTTF}{\text{M}ean \text{T}ime \text{T}o \text{F}ailure}
\newacronym{mttc}{MTTC}{\text{M}ean \text{T}ime \text{T}o \text{C}rash}
\newacronym{noc}{NoC}{\text{N}etwork-on-\text{C}hip}

\newacronym{cots}{COTS}{commercial-off-the-shelf}
\newacronym{nre}{NRE}{\text{N}on \text{R}ecurring \text{E}ngineering}
\newacronym{rtl}{RTL}{\text{R}egister \text{T}ransfer \text{L}evel}
\newacronym{vcu}{VCU}{\text{V}ideo \text{C}odec \text{U}nit}
\newacronym{apu}{APU}{\text{A}pplication \text{P}rocessing \text{U}nit}
\newacronym{gpu}{GPU}{\text{G}raphics \text{P}rocessing \text{U}nit}
\newacronym{rpu}{RPU}{\text{R}eal-time \text{P}rocessing \text{U}nit}
\newacronym{gps}{GPS}{\text{G}lobal \text{P}ositioning \text{S}ystem}
\newacronym{ml}{ML}{\text{M}achine \text{L}earning}
\newacronym{iot}{IoTs}{\text{I}nternet of \text{T}hings}
\newacronym{ic}{ICs}{\text{i}ntegrated \text{c}ircuits}
\newacronym{esl}{ESL} {\text{E}lectronic \text{S}ystem \text{L}evel}
\newacronym{eda}{EDA} {\text{E}lectronic \text{D}esign \text{A}utomation}
\newacronym{clr}{CLR} {\text{C}ross\hyp\text{l}ayer \text{R}eliability}
\newacronym{qos}{QoS} {\text{Q}uality of \text{S}ervice}
\newacronym{hmpsoc}{HMPSoC} {\text{H}eterogeneous \text{M}ulti-\text{P}rocessor \text{S}ystem\hyp on\hyp\text{C}hip}
\newacronym{mpsoc}{MPSoC} {\text{M}ulti-\text{P}rocessor \text{S}ystem\hyp on\hyp\text{C}hip}
\newacronym{soc}{SoC} {\text{S}ystem\hyp \text{o}n\hyp\text{C}hip}
\newacronym{fpga}{FPGA} {\text{F}ield \text{P}rogrammable \text{G}ate \text{A}rray}
\newacronym{dpr}{DPR} {\text{D}ynamic \text{P}artial \text{R}econfiguration}
\newacronym{prr}{PRR} {\text{P}artially \text{R}econfigurable \text{R}egion}
\newacronym{prm}{PRM} {\text{P}artially \text{R}econfigurable \text{M}odule}
\newacronym{pe}{PE} {\text{P}rocessing \text{E}lement}
\newacronym{dse}{DSE} {\text{D}esign \text{S}pace \text{E}xploration}
\newacronym{ga}{GA} {\text{G}enetic \text{A}lgorithms}
\newacronym{bti}{BTI} {\text{B}ias \text{T}emperature \text{I}nstability}
\newacronym{nbti}{NBTI} {\text{N}egative \text{B}ias \text{T}emperature \text{I}nstability}
\newacronym{pbti}{PBTI} {\text{P}ositive \text{B}ias \text{T}emperature \text{I}nstability}
\newacronym{em}{EM} {\text{E}lectro\text{m}igration}
\newacronym{gob}{GOB} {\text{G}ate \text{O}xide \text{B}reakdown}
\newacronym{hci}{HCI} {\text{H}ot \text{C}arrier \text{I}njection}
\newacronym{tddb}{TDDB}{\text{T}ime \text{D}ependent \text{D}ielectric \text{B}reakdown}
\newacronym{seu}{SEU} {\text{S}ingle \text{E}vent \text{U}pset}
\newacronym{ser}{SER} {\text{S}oft \text{E}rror \text{R}ate}
\newacronym{gdb}{GDB} {\text{G}ate \text{D}ielectric \text{B}reakdown}
\newacronym{tmr}{TMR} {\text{T}riple \text{M}odular \text{R}edundancy}
\newacronym{dmr}{DMR} {\text{D}ual \text{M}odular \text{R}edundancy}
\newacronym{ecc}{ECC}{\text{E}rror \text{C}hecking and \text{C}orrecting}
\newacronym{sram}{SRAM}{\text{S}tatic \text{R}andom \text{A}ccess \text{M}emory}
\newacronym{dram}{DRAM}{\text{D}ynamic \text{R}andom \text{A}ccess \text{M}emory}
\newacronym{llc}{LLC}{\text{L}ast \text{L}evel \text{C}ache}
\newacronym{l1}{L1}{\text{L}evel \text{1}}
\newacronym{dimm}{DIMM}{\text{D}ual \text{i}n-line-\text{M}emory \text{M}odule}
\newacronym{snc}{SNC}{\text{S}ingle-\text{N}ibble-error-\text{C}orrecting}
\newacronym{dnd}{DND}{\text{D}ouble-\text{N}ibble-error-\text{D}etecting}
\newacronym{sec}{SEC}{\text{S}ingle-bit-\text{E}rror-\text{C}orrecting}
\newacronym{ded}{DED}{\text{D}ouble-bit-\text{E}rror-\text{D}etecting}
\newacronym{dec}{DEC}{\text{D}ouble-bit-\text{E}rror-\text{C}orrecting}
\newacronym{ted}{TED}{\text{T}riple-bit-\text{E}rror-\text{D}etecting}
\newacronym{ivi}{IVI}{\text{I}nstruction \text{V}ulnerability \text{I}ndex}
\newacronym{fvi}{FVI}{\text{F}unction \text{V}ulnerability \text{I}ndex}
\newacronym{sed}{SED}{\text{S}obel \text{E}dge \text{D}etection}
\newacronym{cnn}{CNN}{\text{C}onvolutional \text{N}eural \text{N}etworks}
\newacronym{dnn}{DNN}{\text{D}eep \text{N}eural \text{N}etworks}
\newacronym{os}{OS}{\text{O}perating \text{S}ystem}
\newacronym{avf}{AVF}{\text{A}rchitectural \text{V}ulnerability \text{F}actor}
\newacronym{milp}{MILP}{\text{M}ixed \text{I}nteger \text{L}inear \text{P}rogramming}
\newacronym{sofr}{SOFR}{\text{S}um-\text{o}f-\text{F}ailure \text{R}ate}
\newacronym{clb}{CLB}{\text{C}onfigurable \text{L}ogic \text{B}locks}
\newacronym{bram}{BRAM}{\text{B}lock \text{RAM}}
\newacronym{dsps}{DSPs}{\text{D}igital \text{S}ignal \text{P}rocessing blocks}
\newacronym{mcts}{MCTS}{\text{M}onte \text{C}arlo \text{T}ree \text{S}earch}
\newacronym{ttp}{TTP} {\text{T}ree \text{T}raversal \text{P}roblem}
\newacronym{fir}{FIR} {\text{F}inite \text{I}mpluse \text{R}esponse}
\newacronym{mtbf}{MTBF}{Mean Time between Failures}
\newacronym{ura}{\textit{uRA}}{User-modulated Run-time Adaptation}
\newacronym{aura}{\textit{AuRA}}{Agent-based User-modulated Run-time Adaptation}
\newacronym{moea}{MOEA}{\text{M}ulti-\text{O}bjective \text{E}volutionary \text{A}lgorithms}
\newacronym{dvfs}{DVFS}{\text{D}ynamic \text{V}oltage and \text{F}requncy \text{S}caling}
 \newacronym{icap}{ICAP}{\text{I}nternal \text{C}onfiguration \text{A}ccess \text{P}ort}
\newacronym{rl}{RL}{\text{R}einforcement \text{L}earning}
\newacronym{pvt}{PVT}{\text{P}rocess, \text{V}oltage, and \text{T}emperature}
\newacronym{nhpp}{NHPP}{\text{N}on-\text{H}omogeneous \text{P}oisson \text{P}rocess}
\newacronym{fit}{FIT}{\text{F}ailures \text{I}n \text{T}ime}
\newacronym{mosfet}{MOSFET}{\text{M}etal \text{O}xide \text{S}emiconductor \text{F}ield \text{E}ffect \text{T}ransistor }
\newacronym{nmos}{NMOS}{\text{N}egative channel \text{M}etal \text{O}xide \text{S}emiconductor}
\newacronym{pmos}{PMOS}{\text{P}ositive channel \text{M}etal \text{O}xide \text{S}emiconductor}
\newacronym{osi}{OSI}{\text{O}pen \text{S}ystems \text{I}nterconnection}
\newacronym{bist}{BIST}{\text{B}uilt-\text{i}n \text{S}elf \text{T}est}
\newacronym{nlp}{NLP}{\text{N}atural \text{L}anguage \text{P}rocessing}
\newacronym{dof}{DoF}{\text{D}egree \text{o}f \text{F}reedom}
\newacronym{lut}{LUT}{Look-Up Table}
\newacronym{ppa}{PPA}{Power-Performance-Area}
\newacronym{behav}{BEHAV}{behavioral accuracy}
\newacronym{mse}{MSE}{Mean Squared Error}
\newacronym{mae}{MAE}{Mean Average Error}
\newacronym{mlp}{MLP}{Multi-Layer Perceptron}
\newacronym{l2}{L2}{Squared Error}
\newacronym{ppf}{PPF}{Pseudo Pareto-front}
\newacronym{vpf}{VPF}{Validated Pareto-front}
\newacronym{rmse}{RMSE}{Root Mean Squared Error}
\newacronym{mac}{MAC}{Multiply-Accumulate}
\newacronym{asic}{ASIC}{Application-specific Integrated Circuit}
\newacronym{gan}{GAN}{Generative Adversarial Network}
\newacronym{ann}{ANN}{Artificial Neural Network}
\newacronym{lpf}{LPF}{Low-pass Filter}
\newacronym{ecg}{ECG}{Electrocardiogram}
\newacronym{shap}{SHAP}{SHapley Additive exPlanations}
\newacronym{bfs}{BFS}{Breadth First Search}
\newacronym{dfs}{DFS}{Depth First Search}
\newacronym{csp}{CSP}{Constraint Satisfaction Problem}
\newacronym{cgp}{CGP}{Cartesian Genetic Programming}
\newacronym{map}{MaP}{Mathematical Programming}
\newacronym{cc}{CC}{Carry-chain}
\newacronym{miqcp}{MIQCP}{Mixed Integer Quadratically Constrained Program}
\newacronym{pr}{PR}{\text{P}olynomial \text{R}egression}
\newcommand{\doctitle}{
    \textit{AxOMaP}: Designing FPGA-based \underline{A}ppro\underline{x}imate Arithmetic \underline{O}perators using \underline{Ma}thematical \underline{P}rogramming 
}
\newcommand{\slidesloc}{https://www.dropbox.com/s/wb0t9sprhwp3zzz/outline.pptx?dl=0}
\newcommand{\titleName}{\textit{AxOMaP}}

\definecolor{add}{rgb}{1, 0, 0}
\definecolor{rev}{rgb}{0, 0, 1}
\definecolor{reg}{rgb}{0, 0, 0}
\colorlet{newcolor}{blue!70!red}
\newcommand{\siva}[1]{\textcolor{black}{#1}}
\newcommand{\salim}[1]{\textcolor{black}{#1}}

\begin{document}
	\glsresetall

\title{
      \doctitle
}

\renewcommand{\shortauthors}{Sahoo et al.}
\author{Siva Satyendra Sahoo}
\affiliation{%
  \institution{Interuniversity Microelectronics Centre (IMEC)}
  \city{Leuven}
  \country{Belgium}
}
\email{Siva.Satyendra.Sahoo@imec.be}
\orcid{0000-0002-2243-5350}

\author{Salim Ullah}
\email{salim.ullah@tu-dresden.de}
\orcid{0000-0002-9774-9522}

\author{Akash Kumar}
\affiliation{%
  \institution{The Chair for Processor Design, Center for Advancing Electronics Dresden (CfAED), Technische Universit{\"a}t Dresden }
  \city{Dresden}
  \country{Germany}
}
\email{akash.kumar@tu-dresden.de}
\orcid{0000-0001-7125-1737}

\renewcommand{\shortauthors}{Sahoo et al.}
\begin{abstract}
\siva{
With the increasing application of machine learning (ML) algorithms in embedded systems, there is a rising necessity to design low-cost computer arithmetic \salim{for these resource-constrained systems}. 
As a result, emerging models of computation\salim{, such as approximate and stochastic computing,} that leverage the inherent error-resilience of such algorithms are being actively explored for implementing ML inference on resource-constrained systems. 
Approximate computing (AxC) aims to provide disproportionate gains in the power, performance, and area (PPA) \salim{of an application} by allowing some level of reduction in \salim{its} behavioral accuracy (BEHAV). 
Using approximate operators (AxOs) for computer arithmetic forms one of the more prevalent methods of implementing AxC. 
\salim{AxOs} provide the additional scope for finer granularity of optimization, compared to only precision scaling of computer arithmetic. 
To this end, the design of platform-specific and cost-efficient approximate operators forms an important research 
\salim{goal}. 
Recently, multiple works have reported the use of AI/ML-based approaches for synthesizing novel FPGA-based AxOs. 
However, most of such works limit the use of AI/ML to designing ML-based surrogate functions that are used during iterative optimization processes. 
To this end, we propose a novel data analysis-driven mathematical programming-based approach to synthesizing approximate operators for FPGAs. 
Specifically, we formulate 
\salim{\textit{mixed integer quadratically constrained programs}}
based on the results of correlation analysis of the characterization data and use the solutions to enable a more directed search approach for evolutionary optimization algorithms. 
Compared to traditional evolutionary algorithms-based optimization, we report up to 21\% improvement in the hypervolume, for joint optimization of PPA and BEHAV, in the design of signed 8-bit multipliers. 
Further, we report up to 27\% better hypervolume than other state-of-the-art approaches to DSE for FPGA-based application-specific AxOs.
}

\end{abstract}



\maketitle

\glsresetall
\section{Introduction}
\label{sec:intro}
\siva{
The last few years have witnessed a steady increase in both the extent and variety of applications using \gls{ml}. 
A large fraction of such applications involves deploying \gls{ml} inference in embedded systems. 
\salim{To this end, the rising complexity of \gls{ml} methods poses a complex problem for resource-constrained embedded systems.}
Consequently, computing paradigms such as \gls{axc} and Stochastic Computing, which reduce the cost of computer arithmetic operations, are being actively explored. 
\gls{axc} denotes a wide variety of methods that can be implemented across the system stack and includes methods such as \salim{loop skipping, precision scaling, and load value approximation ~\cite{venkataramani2015approximate, 10.1109/DAC18074.2021.9586260, mittal2016survey}.} 
For the hardware layer, employing \glspl{axo} for computer arithmetic forms an active area of research. 
\glspl{axo} provide an additional degree of freedom beyond the precision-scaling of arithmetic operations~\salim{\cite{mittal2016survey, shafique2016cross, 10.1145/3566097.3567891}} and, similar to precision scaling, can provide large \gls{ppa} gains by leveraging the applications' \salim{inherent error-resilience} to reduced \gls{behav}. 
Since \gls{ml} inference primarily involves \gls{mac} operations, any \gls{ppa} gains of arithmetic operator implementation can result in considerable system-level \gls{ppa} improvements.
}

\siva{
Designing \glspl{axo} for computer arithmetic usually entails avoiding the implementation of some part of the processing while still ensuring good enough behavioral accuracy. 
This approach is unlike precision scaling, where the operands are scaled to a different bit-width representation scheme. 
Further, precision scaling limits the scope of exploration to just a few possible bit widths. 
Using \glspl{axo}, on the other hand, provides a much finer granularity of implementation and, as a result, can 
\salim{provide} larger scope for application-specific optimizations. 
However, the finer granularity of design and implementation also results in a much larger design space. For instance, as shown in~\cite{evoapprox16}, 
\salim{each combinatorial removal of logic gates in implementing}
an operator can result in a unique \gls{axo}. 
\salim{The consideration of the hardware platform for implementing \glspl{axo} further exacerbates the corresponding \gls{dse} problem.}
For instance, \glspl{axo} designed for \glspl{asic} do not necessarily result in equivalent gains when implemented on \gls{lut}-based \glspl{fpga} \salim{and vice versa~\cite{9072581,9344673, 9218533}.}
}

\begin{figure}[t] 
    \centering
      \scalebox{1}{\includegraphics[width=0.8 \textwidth]{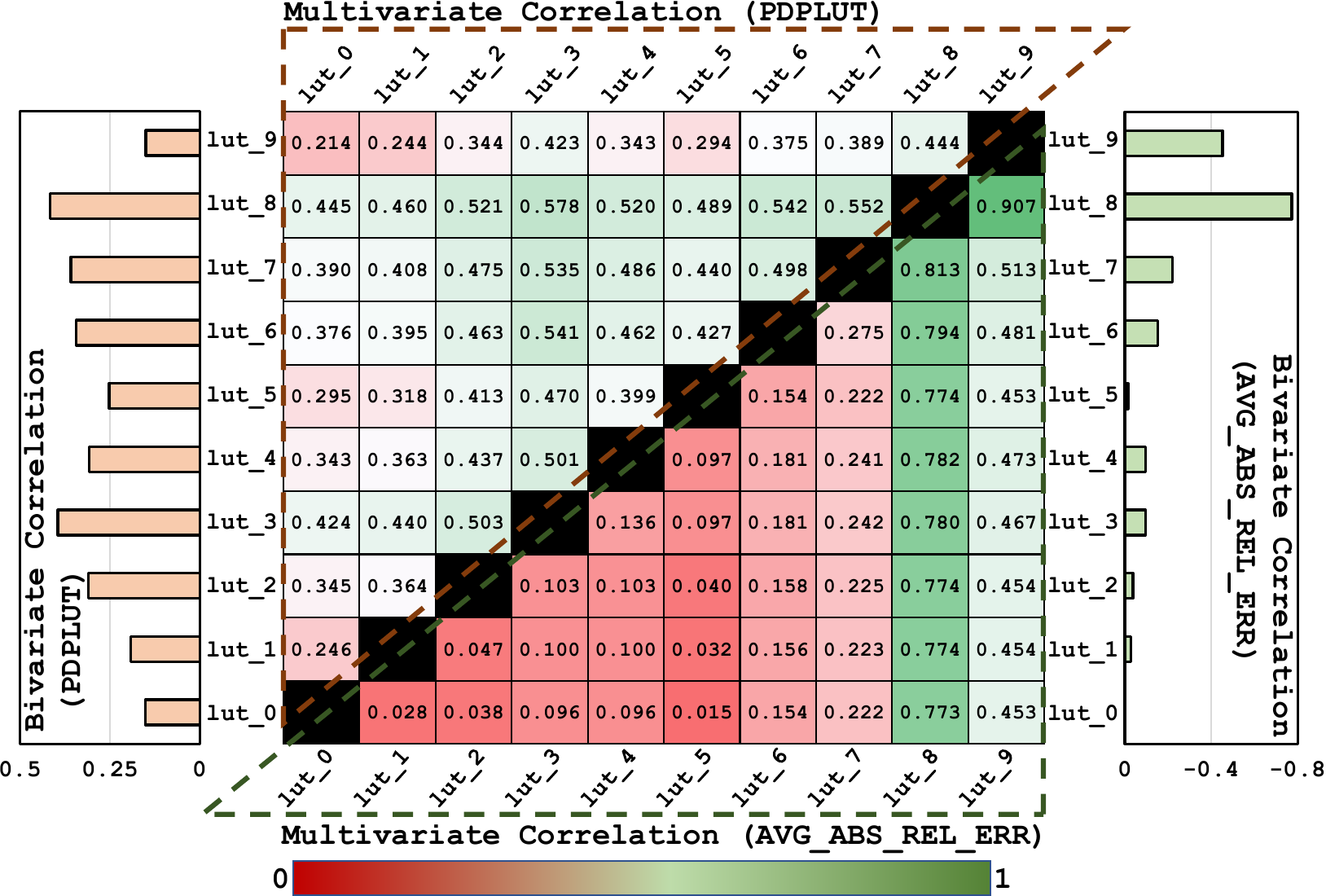}}
  \caption{Correlation between LUT usage and, PDPLUT and AVG\_ABS\_REL\_ERR for approximate signed $4 \times 4$ multiplier designs obtained by selective LUT removal}
  \label{fig:motiv_1} 
\end{figure}
\siva{
Given the diversity of applications being implemented in embedded systems, \glspl{fpga} provide the best \gls{ppa} and time-to-market (cost) trade-offs. 
As a result, \glspl{fpga} are being increasingly used at every scale of computing, from TinyML to compute servers~\cite{prakash2022cfu,8977886}. 
\salim{In this regard, using \glspl{axo} for error-tolerant applications can provide further performance and energy benefits across every scale of computation.}
In our current article, we limit the discussion to the design of \gls{fpga}-based \glspl{axo}. 
The \gls{dse} for such \glspl{axo} has ranged from implementing \gls{asic}-optimized designs on \glspl{fpga} to optimizing the implementation of \glspl{axo} that consider the \gls{lut} and \gls{cc} logic primitives of \glspl{fpga}. 
Recently \gls{ml}-based methods have been proposed for the corresponding \gls{dse} problem. 
However, most of the approaches do not leverage any form of data analysis of the characterization data. 
We present a case study for motivating the use of correlation analysis in generating quantitative metrics that can be used in the \gls{dse} of \gls{fpga}-based \glspl{axo}.
}

\siva{
\autoref{fig:motiv_1} shows the results of the correlation analysis between \gls{lut}-usage and PDPLUT\footnote{PPA: Power $\times$ CPD $\times$ LUT usage} and AVG\_ABS\_REL\_ERR\footnote{BEHAV: Average absolute relative error} metrics in the design of approximate signed $4\times 4$ multipliers. The characterization data corresponds to the 1024 possible designs with the operator model presented in~\cite{ullah2022appaxo}. Accordingly, 10 \glspl{lut} in the implementation of the accurate design were marked for selective removal to generate approximate designs. The bar charts at either end of the figure show the\textit{ bivariate correlation} between each \gls{lut}'s usage\footnote{Using 1/0 to represent the usage/removal of each LUT respectively} and the \salim{corresponding} \gls{ppa} and \gls{behav} metrics. The heatmap in the figure shows the \textit{multivariate correlation} when two \glspl{lut} were considered together \salim{ in the design of an approximate version of the operator}. Any cell in the upper left and bottom right triangle represents the joint correlation of the \glspl{lut} corresponding to the row and column index of the matrix with the \gls{ppa} and \gls{behav} metric, respectively. As can be seen in the figure, there is a non-uniform correlation\salim{, and the analysis provides some indication} of the contribution of each \gls{lut} to the \gls{ppa} and \gls{behav} metrics. \salim{For example, $LUT\_8$ and $LUT\_9$ show a comparatively higher impact on the \gls{behav}. Similarly, for the PDPLUT metric various \glspl{lut} show a comparatively more substantial effect.}
}

\begin{figure}[t] 
    \centering
      \scalebox{1}{\includegraphics[width=0.8 \textwidth]{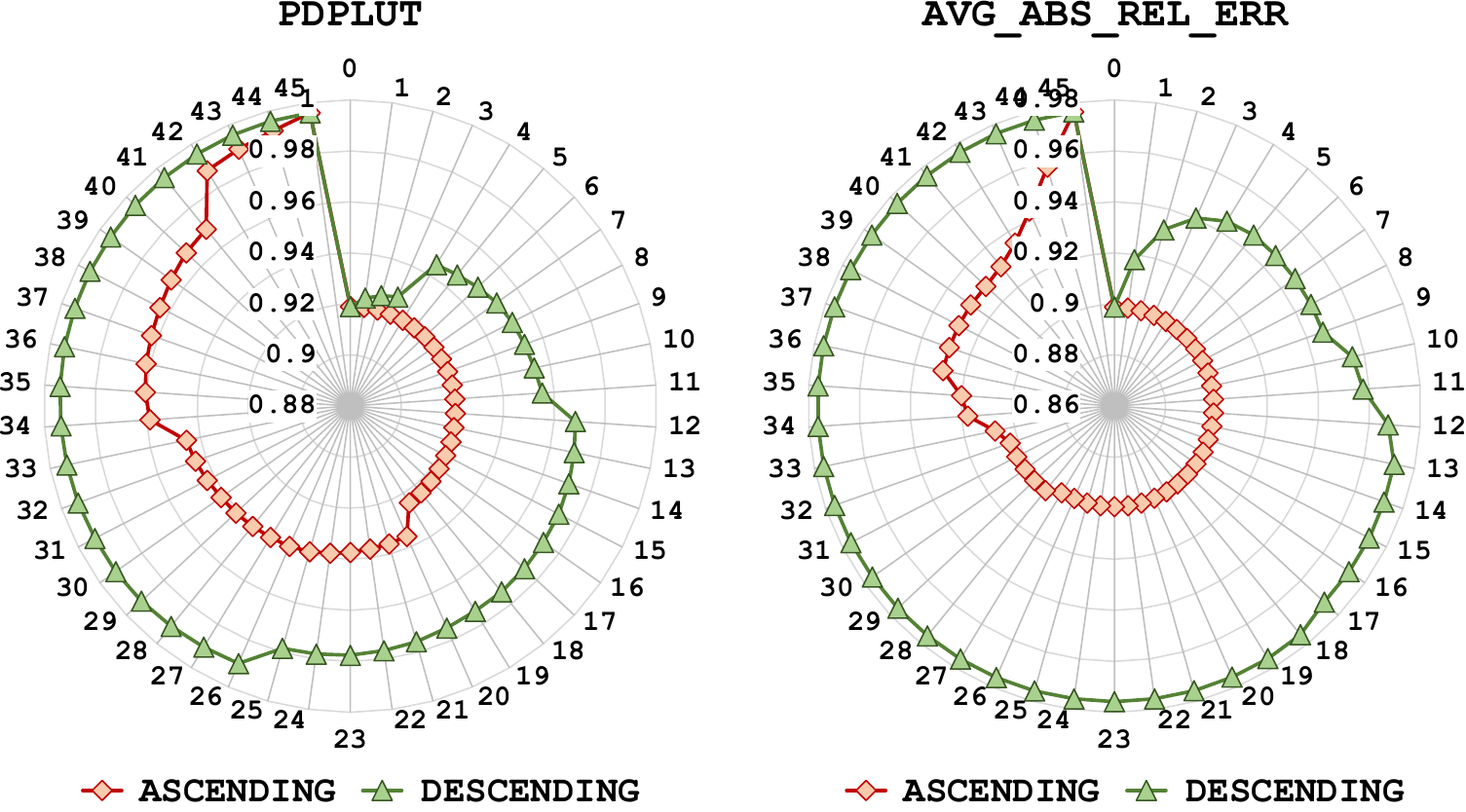}}
  \caption{Comparing the R\textsuperscript{2} score of the polynomial regression models of PDPLUT and AVG\_ABS\_REL\_ERR of approximate signed $4 \times 4$ multipliers with an increasing number of quadratic terms}
  \label{fig:motiv_2} 
\end{figure}

\siva{
We used the multivariate correlation data to rank the joint \glspl{lut} (say lut\_x and lut\_y) and progressively added the product of the corresponding \glspl{lut} (lut\_x $\times$ lut\_y) as an additional feature in the \gls{pr} model for predicting the PPA and BEHAV metrics. \autoref{fig:motiv_2} shows the progression of the R\textsuperscript{2} score (training) of the \gls{pr} model for both metrics. The point $0$ on the radial axis corresponds to a linear regression model, and the other points represent using one additional quadratic feature. The red and green points correspond to the results when the quadratic features were sorted in ascending and descending order, respectively. As evident, adding quadratic features with higher correlation results in a faster increase in the prediction capability of the model. This motivates the use of statistical analysis of characterization data for \gls{dse}.
To this end, we present \titleName, a methodology to leverage correlation analysis in the design of \gls{fpga}-based \glspl{axo}. The related contributions are listed below.\\
\\
\textbf{Contributions:}
\begin{enumerate}[wide, labelwidth=!, labelindent=0pt]
    \item We present a novel approach to formulating the synthesis of \gls{fpga}-based approximate operators as a \gls{map} problem. Specifically, we use the correlation analysis to formulate a \gls{miqcp} problem to generate approximate design configurations.
    \item We present an augmented meta-heuristics optimization method for the corresponding \gls{dse} problem. Specifically, we use results from solving the \glspl{miqcp} to direct a \gls{ga}-based search \salim{on generating} better design points than generic evolutionary optimization approaches.
    \item We evaluate the proposed methodology for both operator-level and application-specific \gls{dse}. Compared to traditional evolutionary algorithms-based optimization, we report up to 21\% improvement in the hypervolume, for joint optimization of \gls{ppa} and \gls{behav}, in the design of signed 8-bit multipliers. Further, we report up to 27\% better hypervolume than other state-of-the-art approaches to \gls{dse} for \gls{fpga}-based application-specific approximate operators. 
\end{enumerate}
}

\siva{
The rest of the article is organized as follows. \autoref{sec:bckRel} presents a brief overview of the requisite background and related works. 
The operator model used in the analysis is presented in~\autoref{sec:op_model}. 
\autoref{sec:propDSE} presents the various components and methods of the proposed \titleName~ methodology. 
The analysis of the results of the experimental evaluation of the proposed methods is presented in~\autoref{sec:expRes}. 
Finally, in~\autoref{sec:conc}, we conclude the article with a summary and a brief discussion of the scope for related future research.
}
\section{Background and Related Works}
\label{sec:bckRel}

\subsection{Approximate Computing}
\salim{\acrfull{axc} paradigm has proven to be a feasible solution to address the ever-increasing computational and memory needs of modern applications ranging from data centers to embedded systems at the edge. Most of these application domains, such as machine learning, computer vision, data mining, and synthesis, offer an inherent error tolerance to the inaccuracies (approximations) in their computations~\cite{venkataramani2015approximate}. \gls{axc} uses this error tolerance to trade the output accuracy of an
application for performance gains. Recent related works have presented various approximation techniques that cover all layers of the system stack~\cite{mittal2016survey, shafique2016cross}. However, for the resource-constrained embedded systems, the architecture and circuit layers have remained the focus of recent related works. For example, employing reduced precision operations is the most commonly employed architectural-level technique~\cite{alemdar2017ternary, 9463531, wang2019bfloat16}. Similarly, using approximate computational units is one of the most commonly used and useful techniques on the circuit layer. Furthermore, as \gls{mac} is one of the main operations in error-tolerant applications, such as \gls{ml}, computer vision, and digital signal processing, most related works have proposed various approximate implementations for adders and multipliers~\cite{10.1145/3566097.3567891, evoapprox16, 9072581, 9344673, ullah2022appaxo, 9233379, 10.1145/3386263.3406907,shafique2015low,ye2013reconfiguration, prabakaran2018demas}.}

\salim{Most of \gls{axo} architectures primarily rely on intentionally introducing approximations for performance gains by truncating portions of some of the calculations or employing inaccurate computations. For example, the authors of~\cite{shafique2015low, ye2013reconfiguration} have proposed to utilize multiple sub-adders to truncate long carry-propagation chains in larger adders. These designs employ multiple preceding bits for each sub-adder to improve the overall accuracy of the larger adder. Similarly, the work presented in~\cite{prabakaran2018demas} has proposed a set of \gls{fpga}-optimized approximate adders by using various carry and sum prediction methods.}

\salim{
The majority of related works have focused on the multiplication operation, owing to its high computational complexity, and proposed various \gls{asic}- and \gls{fpga}-optimized architectures. For instance, the works proposed in~\cite{ko2011design, petra2009truncated} have used various truncation strategies to implement an $\text{M}\times\text{M}$ multiplier by producing an $\text{M}$-bit output. Similarly, other works, such as~\cite{hashemi2015drum}, rely on truncating the input operands to utilize comparatively smaller multipliers to implement a larger multiplier.
Other techniques, such as those proposed in~\cite{10.1145/2966986.2967005} and~\cite{5718826} have utilized functional approximation to implement \gls{asic}-optimized inaccurate $2\times2$ multipliers. The optimized $2\times2$ modules are then utilized to implement larger multipliers. The authors of~\cite{evoapprox16} and~\cite{9233379} have employed Cartesian Genetic Programming (CGP) to present \gls{asic}-optimized approximative adders and multipliers libraries. In their proposed work, they generate approximate operators by executing multiple iterations of the CGP on a set of accurate implementations of an operator. For the CGP iterations, the accurate circuits are represented as strings of numbers, and various approximations of an operator are created using a worst-case error-based objective function. Considering the architectural details of  \glspl{fpga}, the authors of~\cite{9072581}, \cite{9344673}, and~\cite{ullah2018smapproxlib} have proposed various \gls{lut}-level optimization to implement approximate multiplier architecture optimized for Xilinx \glspl{fpga}.
The techniques presented~\cite{9344673} and~\cite{ullah2018smapproxlib} have proposed $4\times4$ approximate multipliers as the building blocks to implement higher-order multipliers. Compared to the \gls{asic}-optimized $2\times2$ designs, \gls{lut}-optimized $4\times4$ multipliers utilize \glspl{lut} more efficiently. The work presented in~\cite{9072581} has focused on implementing signed approximate multipliers. For this purpose, the proposed work uses a radix-4 booth algorithm and employs both functional approximation and truncation of product bits to reduce the total number of utilized \glspl{lut} and carry-propagation chains.  
}

\begin{table}[t]
\centering
\caption{Comparing related works 
}
\label{table:rel}
\small
\def\arraystretch{1}
\resizebox{0.85 \columnwidth}{!}{
\begin{tabular}{@{}cccccccc@{}}
\toprule
Related Works & \cite{9233379,evoapprox16} & \cite{9072581} & \cite{mrazek2019autoax} & \cite{9218533} & \cite{ullah2022appaxo} &  \cite{10.1145/3566097.3567891} & \titleName \\ \midrule
LUT-level Optimization & \xmark & \cmark & \xmark & \xmark & \cmark & \cmark & \cmark \\ \midrule
Application-specific DSE & \xmark & \xmark & \cmark & \cmark & \cmark & \xmark & \cmark \\ \midrule
Automated Search & \cmark & \xmark & \cmark & \xmark & \cmark & \cmark & \cmark \\ \midrule
ML-based Estimation & \xmark & \xmark & \cmark & \cmark & \cmark & \cmark & \cmark \\ \midrule
Iterative Search & \cmark & \cmark & \cmark & \cmark & \cmark & \xmark & \cmark \\ \midrule
Directed Search & \xmark & \xmark & \xmark & \xmark & \xmark & \xmark & \cmark \\ \bottomrule
\end{tabular}
}
\end{table}
\subsection{DSE for Approximate Operators}
\salim{Some recent works, such as~\cite{9218533, mrazek2019autoax, ullah2022appaxo, 10.1145/3566097.3567891}, have employed various techniques to provide libraries of approximate arithmetic operators with varying accuracy-performance trade-offs. These libraries often offer thousands of approximate versions of a single arithmetic operator. For example, the work in~\cite{ullah2022appaxo} can provide up to $2^{36}$ approximate signed $8\times8$ multipliers. Therefore, in many cases, it is rather challenging to identify a feasible implementation of an operator that can satisfy the provided accuracy-performance constraints.
For example, the work presented in~\cite{9072581} has utilized \gls{ga} to find feasible operator implementations for a Gaussian image smoothing application. Similarly, the authors of~\cite{10.1145/3566097.3567891} have used Generative Adversarial Networks (GANs) to find operator configurations meeting provided accuracy-performance constraints.}
\siva{
 \autoref{table:rel} summarizes the different aspects explored by related works for the \gls{dse} of \gls{fpga}-based \glspl{axo}.}

\siva{
While earlier works had focused on implementing \gls{asic}-optimized designs on \glspl{fpga}~\cite{9218533,mrazek2019autoax}, more recent works have focused on synthesizing novel \glspl{axo} that leverage the \gls{lut} and \gls{cc}-based architecture of \glspl{fpga}~\cite{9072581,ullah2022appaxo,10.1145/3566097.3567891,10.1145/3386263.3406907}. 
Across both approaches, recent works have shown increased usage of AI/ML-based methods in the \gls{dse}~\cite{mrazek2019autoax,9218533,ullah2022appaxo,10.1145/3566097.3567891}. 
However, this has been primarily limited to using ML-based estimators of \gls{ppa} and \gls{behav} metrics as surrogate fitness functions in iterative optimization methods. 
Only \cite{10.1145/3566097.3567891} implements a purely ML-based generation of novel \gls{axo} designs. 
However, due to the \textit{open-loop}\footnote{The fitness of newly generated designs is not fed back to the model for iterative improvement} nature of this approach, it can suffer from degraded search results. 
Also, a considerable amount of computations(training) are needed for designing the generative networks. 
Further, none of the works discussed above use any knowledge/information derived from the characterization data to enhance the \gls{dse} search. 
Therefore, we posit that \textit{the results from the data analysis of the characterization data along with low-cost mathematical programming can be used to improve the efficacy of iterative/randomized optimization algorithms}.
To this end, \titleName~ provides a methodology for enabling such improvements in 
evolutionary algorithms-based optimization.
 }   
\clearpage
\section{Operator Model}
\label{sec:op_model}
\begin{figure}[t] 
    \centering
      \scalebox{1}{\includegraphics[width=1 \columnwidth]{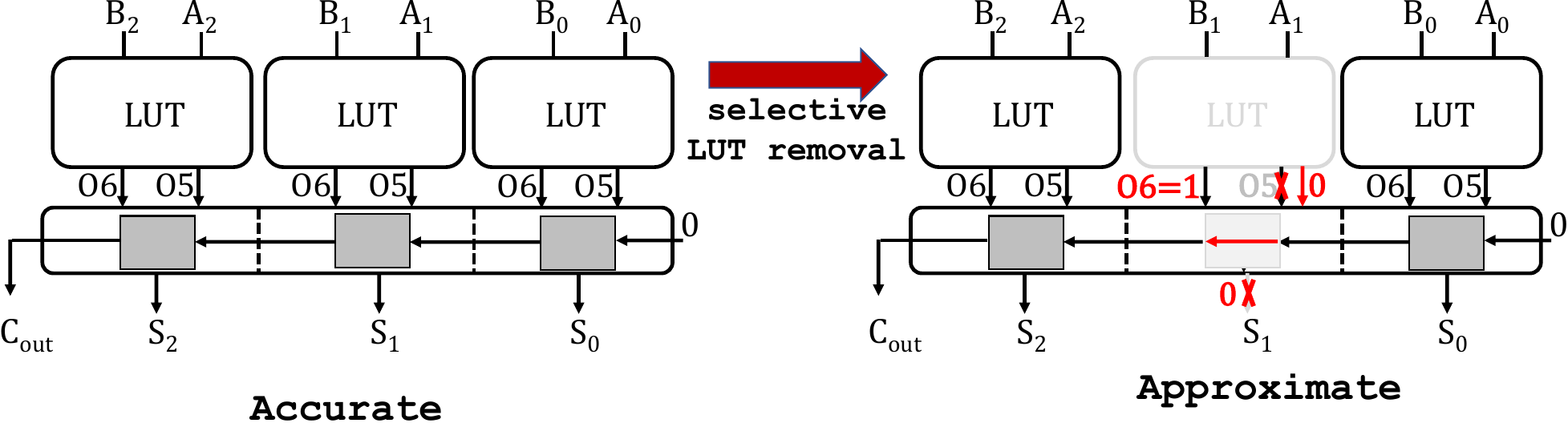}}
  \caption{Approximating a $3-bit$ unsigned adder's FPGA implementation using selective LUT removal}
  \label{fig:op_model} 
\end{figure}

\siva{
The operator model used in the current article is similar to that proposed in~\cite{ullah2022appaxo}. 
Accordingly to the model, any \gls{fpga}-based arithmetic operator can be represented by an ordered tuple: \\ $\mathcal{O}_i(l_0,l_1,..., l_l,..., l_{L-1}), \forall l_l \in \left\{ 0,1 \right\}$ \\
The term $l_l$ represents whether the \gls{lut} corresponding to the operator's accurate implementation is being used or not and $L$ represents the total number of \glspl{lut} of the accurate implementation that may be removed to implement approximation. 
Therefore, $\mathcal{O}_{Ac}(1,1,...,1)$ represents the accurate implementation. 
For instance, the accurate implementation of the  3-bit unsigned adder, shown in~\autoref{fig:op_model}, is represented by the tuple (1,1,1), and 
$\mathcal{O}=\left\{ \mathcal{O}_i \right\}$ represents the set of all possible implementations of the operator. 
Accordingly, the set  $\mathcal{O}$ for the adder shown in the figure is \{ (0,0,0), (0,0,1), (0,1,0),(0,1,1), (1,0,0), (1,0,1), (1,1,0), (1,1,1)\}. 
The approximate implementation in the figure corresponds to the tuple (1,0,1) where the middle \gls{lut} \salim{and the associated inputs are} not used. \salim{Similarly, the output of the associate carry chain cell is also truncated.}
We can abstract any arbitrary operator/application's behavior by a function $\mathcal{S}$. 
So, the operator/application output for a set of inputs can be outlined as shown in~Eq.~\eqref{equ:prob_stmnt_1}. 
The term $Err_{\mathcal{O}_i}$ represents the reduction in the operator/application's behavioral accuracy, \gls{behav}, as a result of using an approximate operator $\mathcal{O}_i$, compared to using the accurate operator $\mathcal{O}_{Ac}$. 
Similarly, the operator/accelerator's \gls{ppa} can be abstracted as a set of functions as shown in~Eq.~\eqref{equ:prob_stmnt_2}.
}

\siva{
\begin{equation}
\label{equ:prob_stmnt_1}
\begin{split}
    Out_{\mathcal{O}_i} = \mathcal{S}(\mathcal{O}_i,Inputs);~ 
    Out_{\mathcal{O}_{Ac}} = \mathcal{S}(\mathcal{O}_{Ac},Inputs) \\
    Err_{\mathcal{O}_i} = Out_{\mathcal{O}_{Ac}} - Out_{\mathcal{O}_i}\\
\end{split}
\end{equation}
}
\siva{
\begin{equation}
\label{equ:prob_stmnt_2}
\begin{split}
    Power~Dissipation: ~~ \mathcal{W}_{\mathcal{O}_i} = \mathcal{H}_W(\mathcal{O}_i,Inputs) \\
    LUT~Utilization: ~~ \mathcal{U}_{\mathcal{O}_i} = \mathcal{H}_U(\mathcal{O}_i) \\
    Critical~Path~Delay: ~~ \mathcal{C}_{\mathcal{O}_i} = \mathcal{H}_C(\mathcal{O}_i) \\
    Power~Delay~Product: ~~PDP_{\mathcal{O}_i} = \mathcal{W}_{\mathcal{O}_i}\times \mathcal{C}_{\mathcal{O}_i} \\
    PDPLUT_{\mathcal{O}_i} = \mathcal{W}_{\mathcal{O}_i}\times \mathcal{U}_{\mathcal{O}_i} \times \mathcal{C}_{\mathcal{O}_i}
\end{split}
\end{equation}
}
\clearpage
\section{\titleName}
\label{sec:propDSE}
\begin{figure}[t]
	\centering
	\scalebox{1}{\includegraphics[width=0.85 \columnwidth]{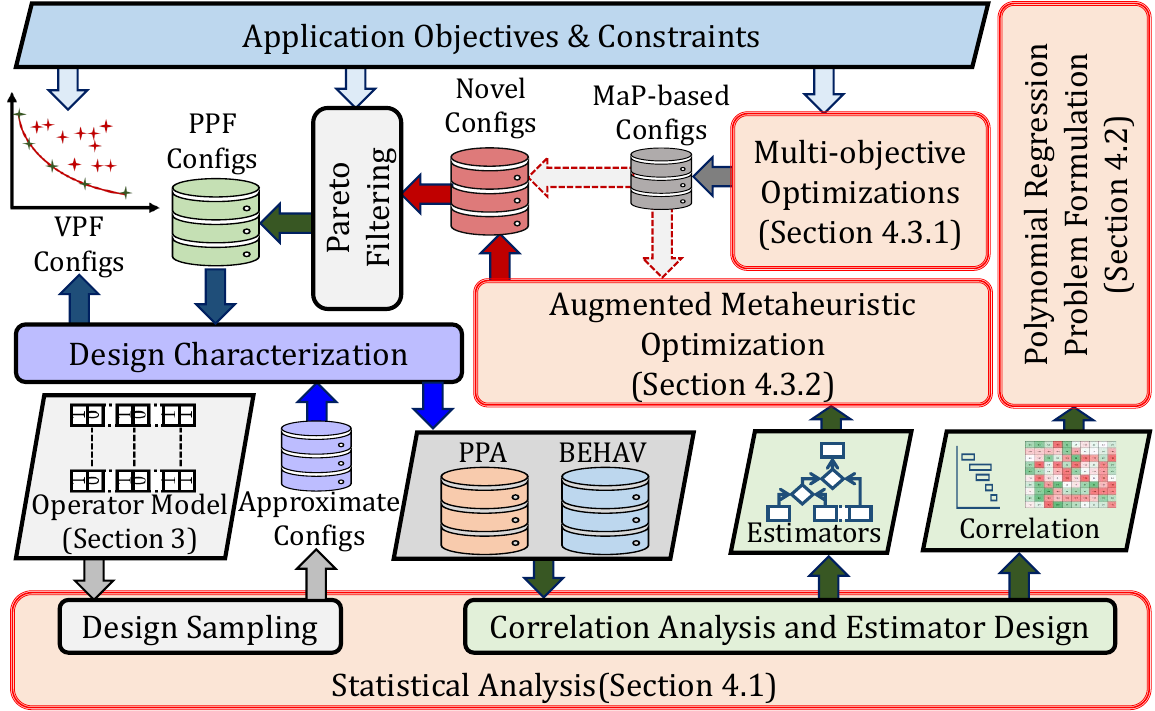}}
	\caption{\titleName~Methodology}
	\label{fig:meth}
\end{figure}
\siva{
The various processes and flow of information in the proposed \titleName~methodology are shown in~\autoref{fig:meth}. 
The proposed contributions are highlighted with the relevant section numbers in the article. 
The statistical analysis involves generating the characterization data used in the correlation analysis and design of the \gls{ml}-based estimators. 
The results of correlation analysis are used in the \gls{pr} design and problem formulation. 
A set of problems are solved for each constrained multi-objective \gls{dse} problem to generate \gls{map}-based approximate configurations. 
These configurations are then passed on as initial solutions for \gls{map}-augmented metaheuristic optimization. 
The resulting solutions are Pareto-filtered using the estimators to generate the \gls{ppf} design configurations which are then characterized (synthesis and implementation) to generate the \gls{vpf} of \gls{fpga}-based \glspl{axo}.
}


\subsection{Statistical Analysis}
\label{subsec:meth_stat}

\subsubsection{Dataset Generation}
\begin{figure}[t]
    \centering
  \subfloat[\label{ppa_dist} PDPLUT]{
    \scalebox{1.0}{
    \includegraphics[width=0.5 \columnwidth]{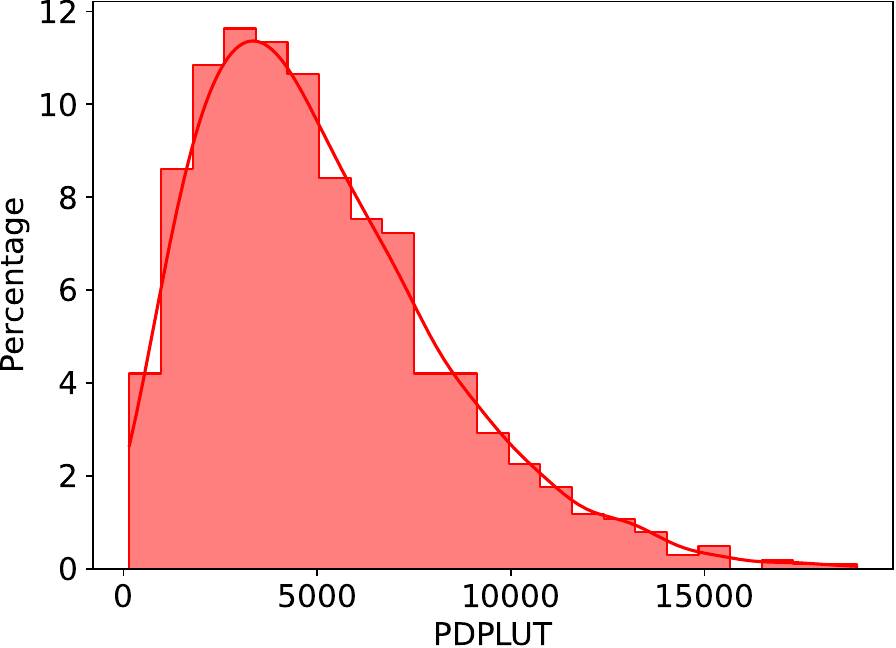}
    }
       }
  \subfloat[\label{behav_dist}AVG\_ABS\_REL\_ERR]{%
        \scalebox{1.0}{\includegraphics[width=0.5 \columnwidth]{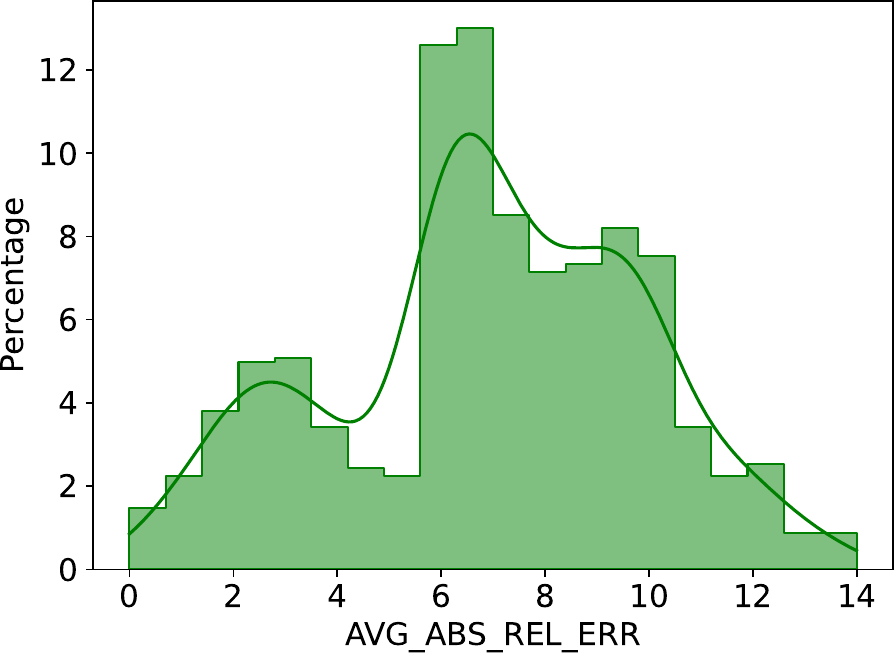}}
        }
\caption{Distribution of the PDPLUT and AVG\_ABS\_REL\_ERR for 1024 approximate signed $4\times 4$ multipliers generated using the operator model proposed in~\cite{ullah2022appaxo}}
  \label{fig:meth_data_dist} 
\end{figure}
\siva{
We used the operator model presented in \textit{AppAxO}~\cite{ullah2022appaxo} to generate a number of approximate designs for characterization and further data analysis. 
While random sampling (with uniform distribution) has been used in related works, we observed a skewed distribution of the resulting metrics. 
For instance, \autoref{fig:meth_data_dist} shows the distribution of the PDPLUT and AVG\_ABS\_REL\_ERR of the set of all possible signed $4\times 4$ multiplier designs that can be generated using the operator model. 
With random sampling in larger operators, the generated points followed a similar distribution, with most points lying within a narrow range for the \gls{ppa} metrics.
Consequently, in order to get a wider range of distribution, we augmented the randomly sampled dataset with design points generated by using different patterns, such as moving windows of consecutive and/or alternating ones and zeros.
}

\subsubsection{Correlation Analysis}
\begin{algorithm}[t]
    \caption{\siva{Computing \textit{multivariate correlation coefficient}}}
    \label{alg:mulCorr}
\siva{
\begin{algorithmic}[1]
    \REQUIRE {$\mathcal{D}$, $x,~y$, $\mathcal{M}$}\\
    Metric under consideration: $\mathcal{M}$ \\
    Characterization dataset: $\mathcal{D}$: $ \left \{ (\mathcal{O}_i,\mathcal{M}_i) \right \}$ \\
    Arbitrary approximate configuration: $\mathcal{O}_i$ \\
    Selected LUTs: $x,~y$ \\
    \STATE {Filter columns based on selected LUTs: $x, ~y$}
    \STATE {Train Regression model: $\mathcal{M} = c_o + c_1 \times l_x + c_2 \times l_y$}
    \STATE {Compute R\textsuperscript{2} score of the regression model: $r^2$}
    \STATE {Return $r = \sqrt{r^2}$}
    \end{algorithmic}
}
\end{algorithm}
\siva{
The results of the correlation analysis for approximate designs of signed $4\times 4$ multipliers were presented in~\autoref{fig:motiv_1}. 
While the bivariate correlation data was generated by measuring the Pearson correlation coefficient, the multivariate correlation was generated using \autoref{alg:mulCorr}.
The multivariate correlation essentially represents the capability of the selected independent variables (features/\glspl{lut}) to predict the dependent variable (PPA/BEHAV metric) in a linear regression model. 
Therefore, as shown in \autoref{alg:mulCorr}, the square root of the R\textsuperscript{2} score of the corresponding linear regression model, with the selected variables, is used to represent the multivariate correlation coefficient.
}

\subsubsection{Estimator Design}
\siva{
The current article focuses on using the characterization data more efficiently than just being limited to designing surrogate factions for estimating \gls{ppa} and \gls{behav} metrics. 
Therefore, we used AutoML~\cite{mljar} to explore across \gls{ml} models and their respective hyperparameters to determine the best estimator for each metric. 
}


\subsection{Problem formulation}
\siva{
The \gls{map} aspect of \titleName~involves using \acrfull{pr} to generate \gls{milp} and \acrfull{miqcp} problems.
}

\subsubsection{Variables}
\siva{
According to the operator model discussed in \autoref{sec:op_model}, the \gls{lut} usage determines the approximate design completely. 
Therefore, the \gls{lut} usage variables, $l_i, \forall i \in \left\{ 0,1, 2, ..., L-1 \right\}$ forms the set of decision variables of the problem.
Further, since we consider multi-objective optimization problems, i.e. minimizing one \gls{ppa} metric and one \gls{behav} metric, we use two support variables to represent the metrics, $v_{ppa}$ and $v_{behav}$. 
The expression for these support variables is based on the polynomial regression results. 
Let us denote the set of \gls{pr} coefficients as $p_{i,j}$ and $b_{i,j}$ where $p_{i,j}$ and $b_{i,j}$ represent the coefficients of the terms of the \gls{pr} model of \gls{ppa} and \gls{behav} respectively. 
For a linear model (and the corresponding \gls{milp} problem) we constraint $i=j$. 
Since we consider only binary decision variables, $l_i \times l_j$ is equivalent to $l_i$ for $i=j$. 
The corresponding expression for $v_{ppa}$ and $v_{behav}$ are shown in~Eq.~\eqref{eq:sup_milp}. 
If we allow quadratic terms in the \gls{pr} models, for \gls{miqcp} problems, we use the relationship shown in~Eq.~\eqref{eq:sup_miqcp}.
}

\siva{
\begin{equation}
\label{eq:sup_milp}
\begin{split}
    v_{ppa}=\sum_{i=0}^{L-1} p_{i,i}\times l_i \\
    v_{behav}=\sum_{i=0}^{L-1} b_{i,i}\times l_i
\end{split}
\end{equation}
}

\siva{
\begin{equation}
\label{eq:sup_miqcp}
\begin{split}
    v_{ppa}=\sum_{(i,j)\forall i,j \in \left \{0,1,...L-1\right \}} p_{i,j}\times l_i~l_j \\
v_{behav}=\sum_{(i,j)\forall i,j \in \left \{0,1,...L-1\right \}} b_{i,j}\times l_i~l_j
\end{split}
\end{equation}
}

\subsubsection{Constraints}
\siva{
The set of constraints is shown in~Eq.~\eqref{eq:sup_const}.
The first set of constraints corresponds to the binary nature of the decision variables. 
The second set of constraints corresponds to the constraints set on the \gls{ppa} and \gls{behav} metrics. 
}

\siva{
\begin{equation}
\label{eq:sup_const}
\begin{split}
    l_i \in  \left  \{0,1\right \} &~\forall ~i \in \left \{0,1,..., L-1\right \}\\
    v_{behav} \leq  max_{behav}~&;~~ v_{ppa} \leq  max_{ppa}
\end{split}
\end{equation}
}

\subsubsection{Objective}
\siva{
Eq.~\eqref{eq:obj_1} shows the generic representation of the multi-objective optimization problem.
Specific formulations to generate a solution pool are discussed next.
}

\siva{
\begin{equation}
\label{eq:obj_1}
\begin{split}
    \underset{\mathcal{O}_i \in \mathcal{O}}{\text{minimize}} ({BEHAV}_{\mathcal{O}_i},{PPA}_{\mathcal{O}_i}) 
    \\
    s.t. ~ {BEHAV}_{\mathcal{O}_i} \leq B_{MAX} ~~ and ~~ {PPA}_{\mathcal{O}_i} \leq P_{MAX}
\end{split}
\end{equation}
}

\subsection{Generating Pareto-front Designs}

\subsubsection{Generating Solution Pool}
\siva{
In order to generate a pool of solutions for each optimization problem, we use a weighted sum of the BEHAV and PPA metrics to formulate multiple objective functions. Eq.~\eqref{eq:obj_2} shows the objective function used for each \gls{map} problem. We sweep the value of $wt_{B}$ from 0 to 1 in increments of 0.05, to generate nearly 20 optimization problems. Further, we also vary the number of quadratic features in the support variable expressions of~Eq.~\eqref{eq:sup_miqcp} to generate multiple \gls{map} problems. 
}
\siva{
We used the multivariate correlation coefficients to rank the product features and incrementally add one additional feature to Eq.~\eqref{eq:sup_milp} to generate a new \gls{map} problem. 
The corner cases of this approach are the MILP where no quadratic features are used and the case where all possible quadratic terms are included in Eq.~\eqref{eq:sup_miqcp}.
For approximating the signed $4 \times 4$ multiplier, with 10 removable \glspl{lut}, it results in 45 quadratic terms. Similarly,  for a signed $8 \times 8$ multiplier with 36 removable \glspl{lut} it amounts to 630 quadratic terms. Hence, for larger designs using all possible quadratic terns can result in large \gls{map} problems, and the correlation analysis can be used to select a subset of the possible terms for the problem formulation.
}

\siva{
We used the maximum PPA and BEHAV metrics reported in the training dataset, denoted by $P_{MAX}$ and $B_{MAX}$ respectively, to formulate different constrained optimization problems. Specifically, we use  constraint scaling factor, $const\_sf$, values of 0.2, 0.5, 0.8, 1.0, 1.2, and 1.5 to determine the $max_{ppa}$ and $max_{behav}$ of~Eq.~\eqref{eq:sup_const}, as shown in~Eq.~\eqref{eq:const_scaling}. Each value of $const\_sf$, and the resulting constrained optimization problem solutions indicate the efficacy of the \gls{dse} approach of the solver under different degrees of tight constraints.
}

\siva{
\begin{equation}
\label{eq:obj_2}
\begin{split}
    \underset{\mathcal{O}_i \in \mathcal{O}}{\text{minimize}} (wt_B \times {BEHAV}_{\mathcal{O}_i} + (1-wt_B){PPA}_{\mathcal{O}_i}) 
\end{split}
\end{equation}
}
\siva{
\begin{equation}
\label{eq:const_scaling}
\begin{split}
    max_{ppa}=const\_sf \times P_{MAX}\\
    max_{behav}=const\_sf \times B_{MAX}
\end{split}
\end{equation}
}

\subsubsection{Augmented Metaheuristic Optimization}
\begin{figure}[t]
	\centering
	\scalebox{1}{\includegraphics[width=0.75 \columnwidth]{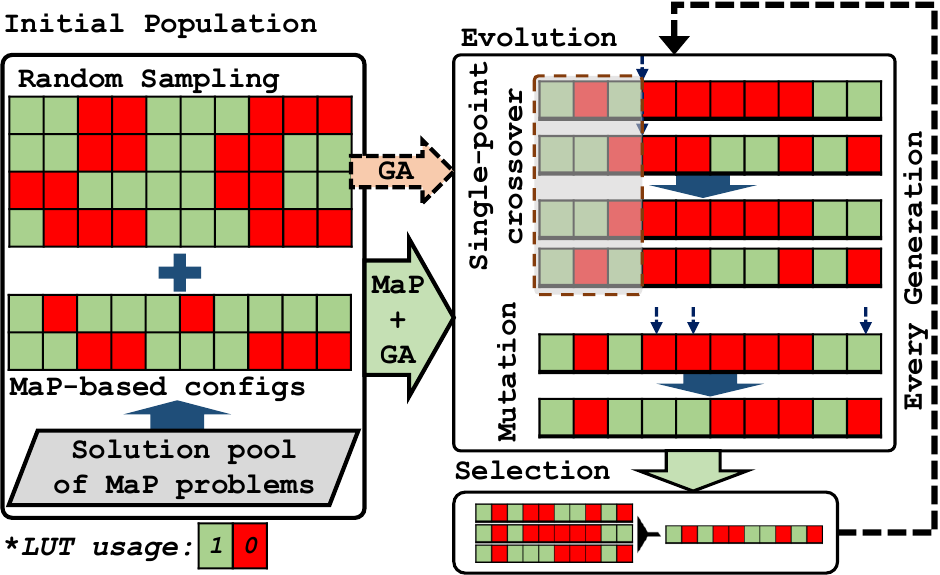}}
	\caption{Genetic Algorithms-based multi-objective DSE approaches explored in \titleName~for synthesizing approximate operators}
	\label{fig:dse_meth}
\end{figure}
\siva{
We use \gls{ga} as an example of a metaheuristics-based solver for the \gls{dse} of \gls{fpga}-based \glspl{axo}. 
\gls{ga} involves generating an initial population of sample solutions and selecting the population for the next generation from a set of solutions obtained by crossover and mutation of the current population. 
We used tournament selection and single-point crossover with a maximum of 250 generations for each experiment.
}
\siva{
As shown in~\autoref{fig:dse_meth}, the problem-agnostic \gls{ga} method involves using random sampling to generate the initial population. We also use an augmented approach where we use the solutions pool generated by solving the \gls{map} problems as the initial population, in addition to the random initial configurations. 
This allows us to direct the search towards pseudo-optimal solutions faster using \gls{map}.
}

\clearpage
\section{Experiments and Results}
\label{sec:expRes}
\subsection{Experiment Setup}

\siva{
\autoref{table:exp_app} shows the designs used for the experimental evaluation along with the relevant PPA and BEHAV metric used in each case.
We use the implementation of approximate signed $8 \times 8$ multipliers, both as application-agnostic (OP) and as application-specific,  designs for the experiments.  
These approximation configurations are implemented in VHDL and synthesized for the $7VX330T$ device of the Virtex-7 family using Xilinx Vivado 19.2. 
The dynamic power is computed by recording the dynamic switching activity for all possible input combinations of the multiplier configurations. 
For this purpose, we have used Vivado Simulator and Power Analyzer tools.
The \gls{map} and \gls{dse} methods are implemented in Python, utilizing packages such as DEAP~\cite{DEAP_JMLR2012}, PyGMO~\cite{Biscani2020} and Scikit~\cite{scikit-learn} among others.
}

\begin{table}[t]
\centering
\caption{Designs used for experiments}
\label{table:exp_app}
\small
\def\arraystretch{1.2}
\resizebox{0.9 \columnwidth}{!}{
\begin{tabular}{|c|c|c|c|c|}
\hline
Design & Description & Accelerator & PPA metric & BEHAV Metric\\ \hline
OP & Signed $8 \times 8$ Multiplier & OP & PDPLUT & AVG\_ABS\_REL\_ERR \\ \hline
ECG & \begin{tabular}[c]{@{}c@{}}Low-pass filter in ECG  \\ peak detection\end{tabular} & 1D Conv & PDPLUT & Peak detection error \\ \hline
MNIST & \begin{tabular}[c]{@{}c@{}}Last dense layer in MNIST\\ digit recognition\end{tabular} & GEMV & PDPLUT & Classification error \\ \hline
GAUSS & \begin{tabular}[c]{@{}c@{}}Gaussian smoothing\\ using 2D convolution\end{tabular} & 2D Conv & PDPLUT & \begin{tabular}[c]{@{}c@{}}Average reduction \\ in PSNR\end{tabular} \\ \hline
\end{tabular}
}
\end{table}


\subsection{
Statistical Analysis
}
\subsubsection{Characterization Dataset}
\begin{figure}[t]
    \centering
  \subfloat[\label{ppa_dist1} PDPLUT]{
    \scalebox{1.0}{
    \includegraphics[width=0.5 \columnwidth]{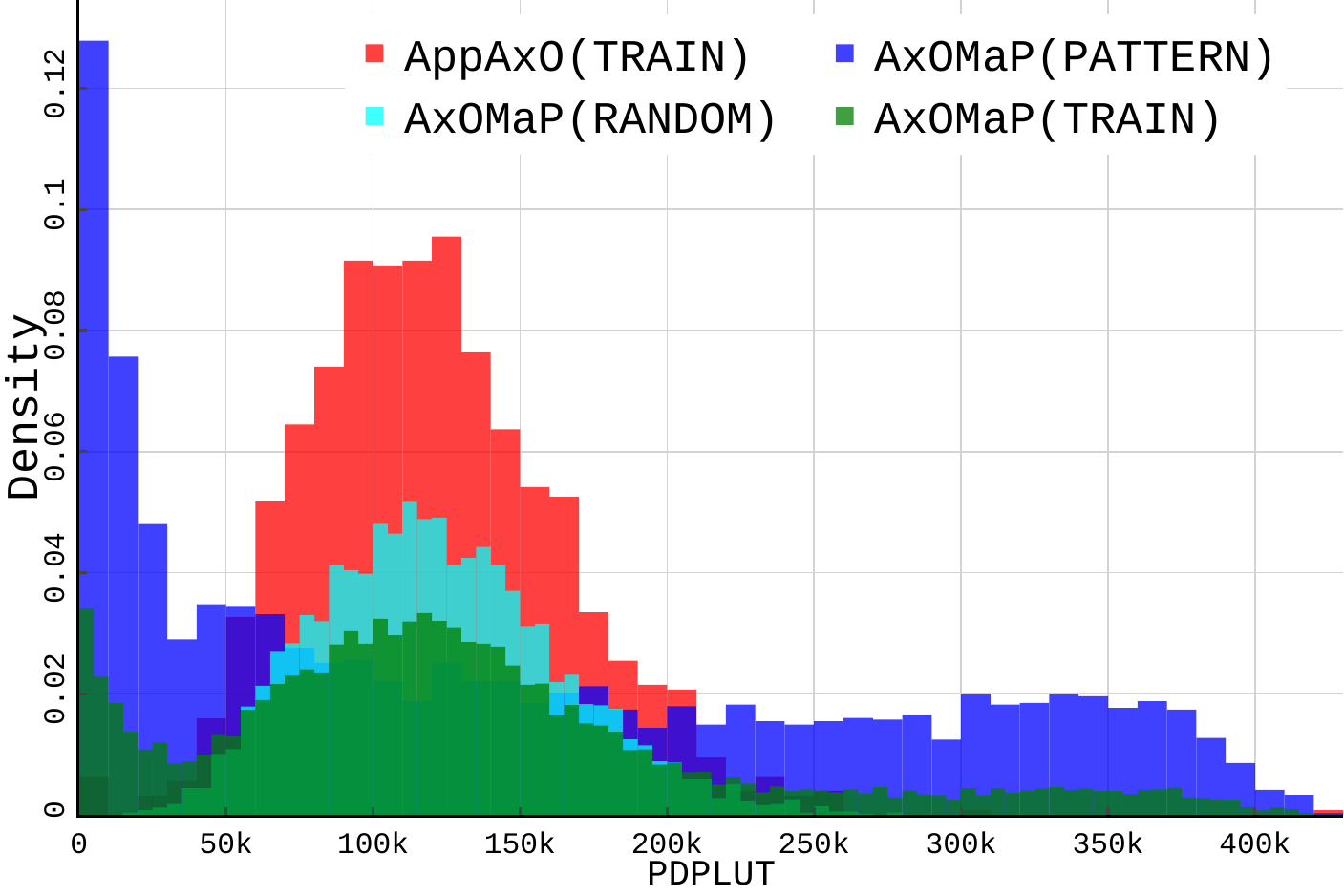}
    }
       }
  \subfloat[\label{behav_dist1}AVG\_ABS\_REL\_ERR]{%
        \scalebox{1.0}{\includegraphics[width=0.5 \columnwidth]{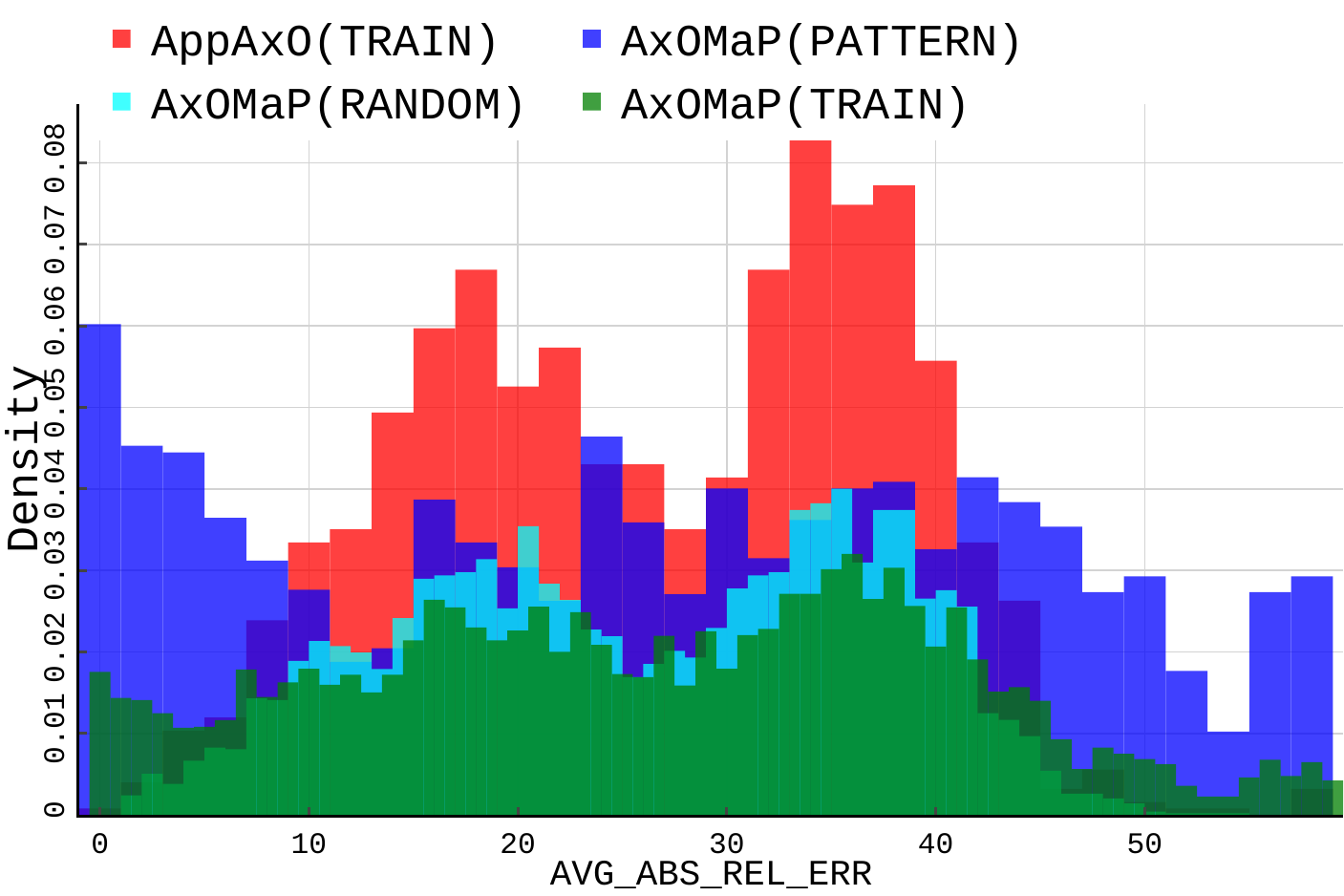}}
        }
\caption{Comparison of the distribution of PDPLUT and AVG\_ABS\_REL\_ERR for approximate signed $8\times 8$ multipliers obtained by different methods of generating the characterization dataset}
  \label{fig:exp_data_dist_1} 
\end{figure}

\begin{figure}[t]
    \centering
  \subfloat[\label{ppa_dist2} PPA]{
    \scalebox{1.0}{
    \includegraphics[width=0.46 \columnwidth]{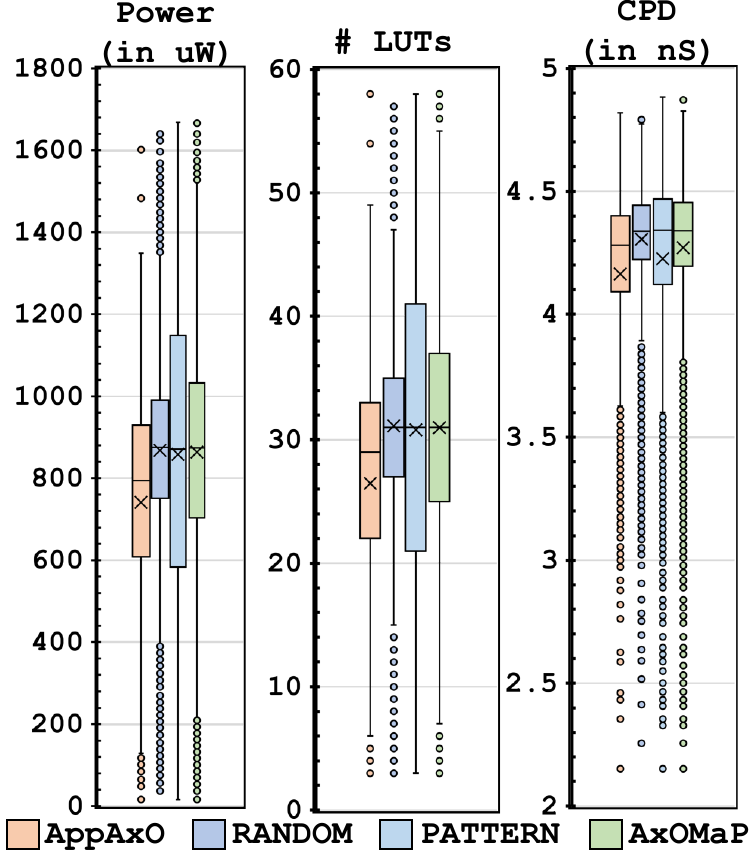}
    }
       }
\hspace{0.01\textwidth}
  \subfloat[\label{behav_dist2}BEHAV]{%
        \scalebox{1.0}{\includegraphics[width=0.42 \columnwidth]{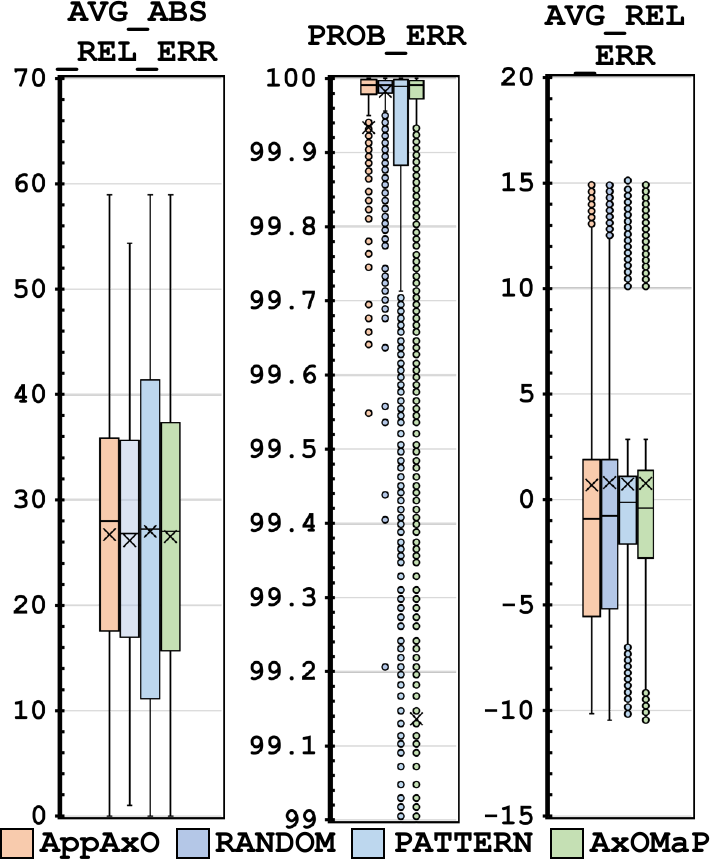}}
        }
\caption{Comparison of the distribution of multiple PPA and BEHAV metrics for approximate signed $8\times 8$ multiplier for \textit{AppAxO}~\cite{ullah2022appaxo} and \titleName~training datasets}
  \label{fig:exp_data_dist_2} 
\end{figure}

\siva{
\autoref{fig:exp_data_dist_1} shows the distribution of PDPLUT and AVG\_ABS\_REL\_ERR of approximate signed $8 \times 8$ multiplier designs generated using different approaches. 
The AppAxO(TRAIN)~\cite{ullah2022appaxo} and the \titleName(RANDOM) datasets show similar distributions as they follow primarily a uniform distribution-based random sampling. 
The \titleName(PATTERN) refers to the dataset generated as described in~Section~\ref{subsec:meth_stat}. 
Evidently, it results in a more balanced distribution across a wider range of the PPA metrics. 
The \titleName(TRAIN) refers to the dataset used for further analysis and modeling and combines the RANDOM and PATTERN datasets. 
\autoref{fig:exp_data_dist_2} shows the box plots of various PPA and BEHAV metrics for the \textit{AppAxO} and \titleName~datasets. 
The \titleName~dataset exhibits a wider range across all metrics. 
Specifically, for the \textit{Probability of Error} (PROB\_ERR) we\ observe many design points under 99.4\%, while the \textit{AppAxO} dataset has none. 
While some of the improvements result from the larger dataset (10,650 compared to 2000), most of the improvements result from the PATTERN dataset, which shows a wider range across most metrics.
}
\subsubsection{Correlation Analysis}
\begin{figure}[htb!] 
    \centering
  \subfloat[\label{exp_correl_bivar} Bivariate Correlation]{
    \scalebox{1.0}{
    \includegraphics[width=0.5 \columnwidth]{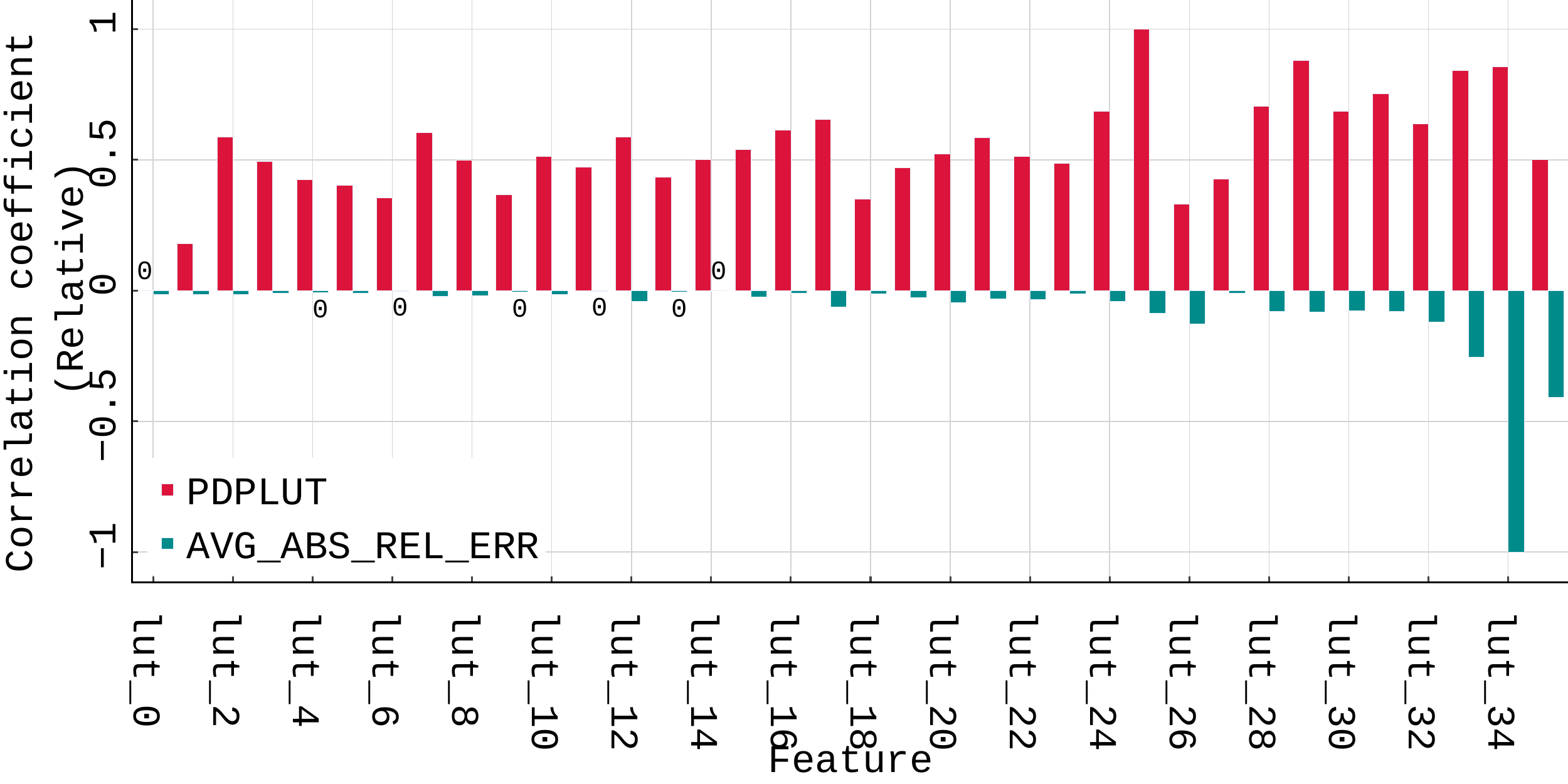}
    }
       }
  \subfloat[\label{exp_correl_multi}Multivariate Correlation]{%
        \scalebox{1.0}{\includegraphics[width=0.5 \columnwidth]{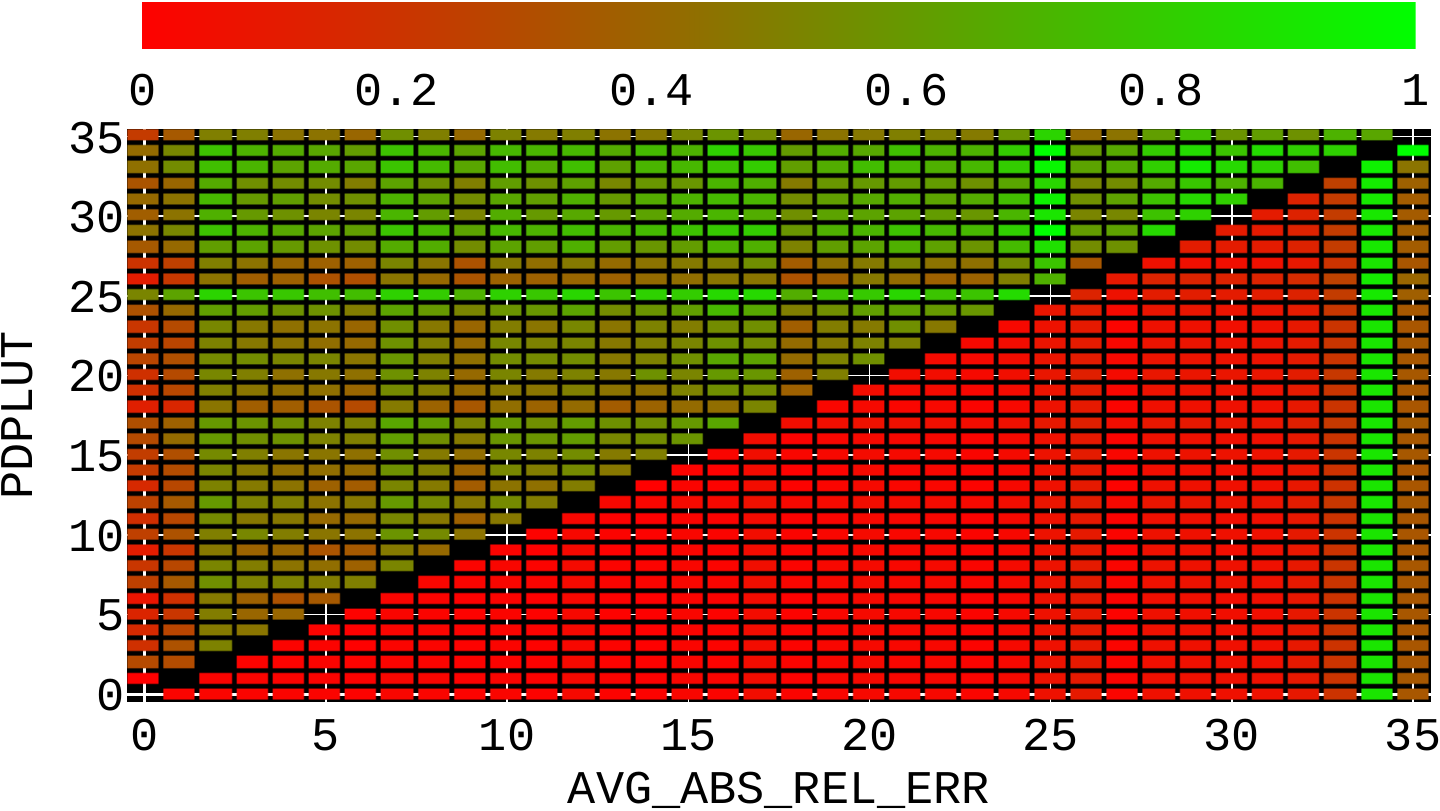}}
        }
\caption{Results of bivariate and multivariate correlation analysis of the training dataset of signed $8\times 8$ approximate multipliers}
  \label{fig:exp_correl} 
\end{figure}
\siva{
\autoref{fig:exp_correl} shows the results of the correlation analysis of the characterized dataset for approximate signed $8 \times 8$ multipliers. \autoref{fig:exp_correl}(a) shows the relative bivariate correlation coefficients, of each of the 36 removable LUTs,  with PDPLUT and AVG\_ABS\_REL\_ERR. Similar to the data shown for the signed $4 \times 4$ multipliers presented in~\autoref{fig:motiv_1}, the correlation coefficients for PDPLUT show a wider distribution across LUTs' usage than AVG\_ABS\_REL\_ERR, where a few LUTs show much larger  correlation coefficient values than others.
This imbalance is also reflected in the multivariate correlation data shown in  \autoref{fig:exp_correl}(b) where the LUTs with higher bivariate correlation coefficients show higher multivariate correlation when analyzed alongside other LUTs.
}
\subsubsection{Estimator Design}
\siva{
\autoref{table:exp_est_model} shows the results of AutoML for different PPA and BEHAV metrics for the approximate signed $8 \times 8$ multiplier characterization dataset. 
Since the features (LUT usage) are primarily categorical than numerical, \gls{ml} algorithms that are tailored for categorical features, such as CatBoost, show better performance in the regression models. 
Almost all models (except for CPD) show R\textsuperscript{2} scores of more than 0.9 across training and testing datasets. 
Also, metrics that are products of other metrics, such as PDP and PDPLUT, exhibit much higher \gls{mse} and \gls{mae} values than others.
}
\begin{table}[htb!]
\centering
\caption{Regression model metrics for various PPA and BEHAV metrics of approximate implementations of signed $8\times 8$ multipliers}
\label{table:exp_est_model}
\small
\def\arraystretch{1.1}
\resizebox{0.8 \columnwidth}{!}{
\begin{tabular}{|cc|cc|cc|cc|}
\hline
\multicolumn{2}{|c|}{Model Metric}                        & \multicolumn{2}{c|}{MSE}           & \multicolumn{2}{c|}{MAE}           & \multicolumn{2}{c|}{R2 Score}      \\ \hline
\multicolumn{1}{|c|}{Design Metric}      & Selected Model & \multicolumn{1}{c|}{TRAIN} & TEST  & \multicolumn{1}{c|}{TRAIN} & TEST  & \multicolumn{1}{c|}{TRAIN} & TEST  \\ \hline
\multicolumn{1}{|c|}{AVG\_ABS\_ERR}      & CatBoost       & \multicolumn{1}{c|}{396}   & 5582  & \multicolumn{1}{c|}{14}    & 54    & \multicolumn{1}{c|}{0.999} & 0.999 \\ \hline
\multicolumn{1}{|c|}{AVG\_ABS\_REL\_ERR} & CatBoost       & \multicolumn{1}{c|}{0.015} & 0.03  & \multicolumn{1}{c|}{0.088} & 0.125 & \multicolumn{1}{c|}{0.999} & 0.999 \\ \hline
\multicolumn{1}{|c|}{PROB\_ERR}          & LightGBM       & \multicolumn{1}{c|}{0.289} & 1.505 & \multicolumn{1}{c|}{0.122} & 0.293 & \multicolumn{1}{c|}{0.989} & 0.929 \\ \hline
\multicolumn{1}{|c|}{POWER}              & CatBoost       & \multicolumn{1}{c|}{206}   & 768   & \multicolumn{1}{c|}{11}    & 22    & \multicolumn{1}{c|}{0.998} & 0.992 \\ \hline
\multicolumn{1}{|c|}{CPD}                & LightGBM       & \multicolumn{1}{c|}{0.012} & 0.018 & \multicolumn{1}{c|}{0.086} & 0.104 & \multicolumn{1}{c|}{0.877} & 0.819 \\ \hline
\multicolumn{1}{|c|}{LUTs}               & CatBoost       & \multicolumn{1}{c|}{0.04}  & 0.108 & \multicolumn{1}{c|}{0.158} & 0.257 & \multicolumn{1}{c|}{0.999} & 0.999 \\ \hline
\multicolumn{1}{|c|}{PDP}                & CatBoost       & \multicolumn{1}{c|}{1596}  & 6797  & \multicolumn{1}{c|}{30}    & 64    & \multicolumn{1}{c|}{0.999} & 0.997 \\ \hline
\multicolumn{1}{|c|}{PDPLUT}             & CatBoost       & \multicolumn{1}{c|}{6.6E6} & 1.4E6 & \multicolumn{1}{c|}{1951}  & 2909  & \multicolumn{1}{c|}{0.999} & 0.998 \\ \hline
\end{tabular}
}
\end{table}


\subsection{
Design Space Exploration
}
\subsubsection{\gls{map}-based DSE}
\begin{figure}[t]
	\centering
	\scalebox{1}{\includegraphics[width=0.8 \textwidth]{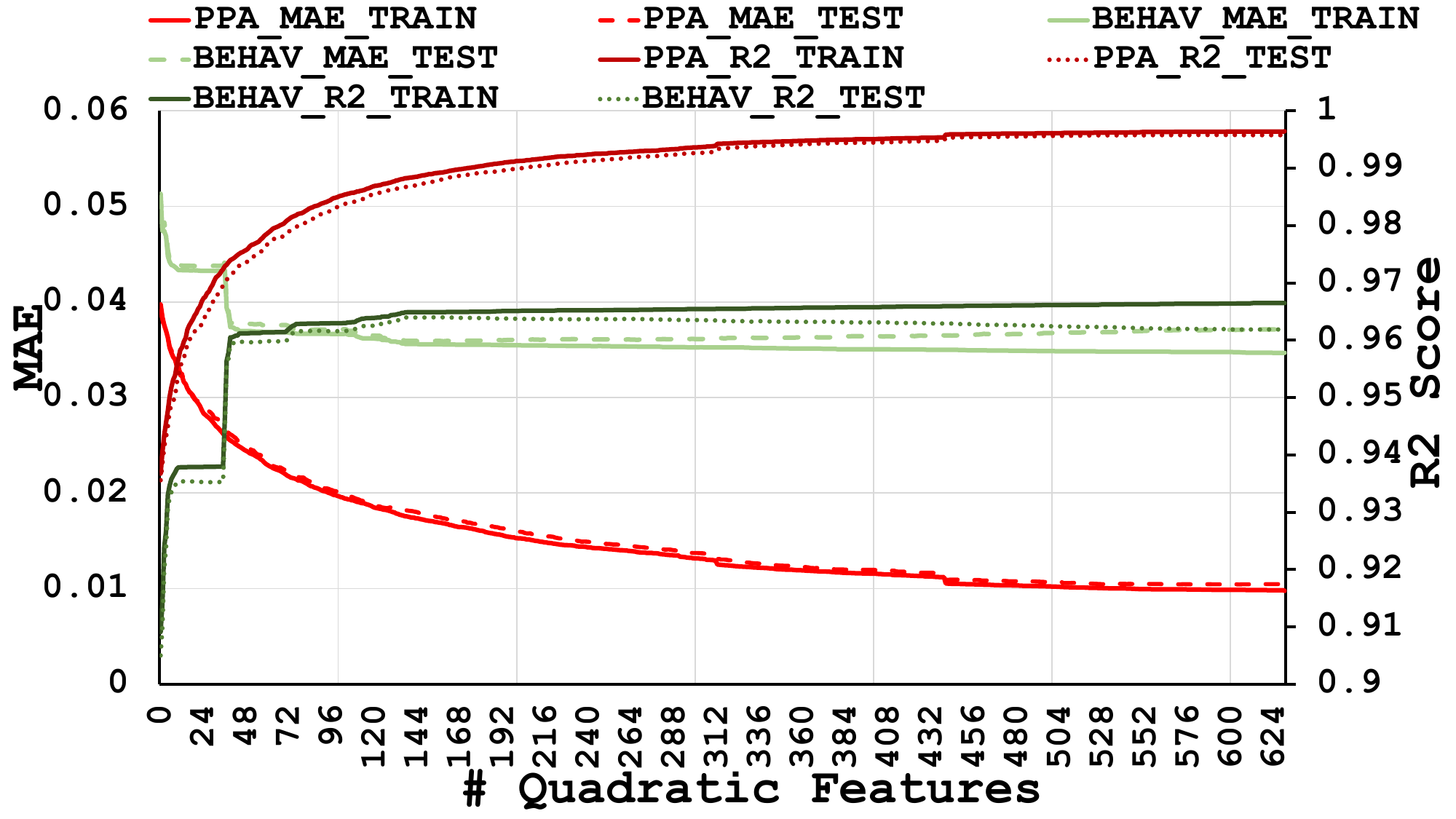}}
	\caption{Accuracy metrics of polynomial regression models with an increasing number of quadratic terms. \textit{MinMaxScaling} was used before generating the models. PPA and BEHAV refer to  PDPLUT and AVG\_ABS\_REL\_ERR respectively}
	\label{fig:exp_map_models}
\end{figure}
\begin{figure}[t]
	\centering
	\scalebox{1}{\includegraphics[width=0.8 \textwidth]{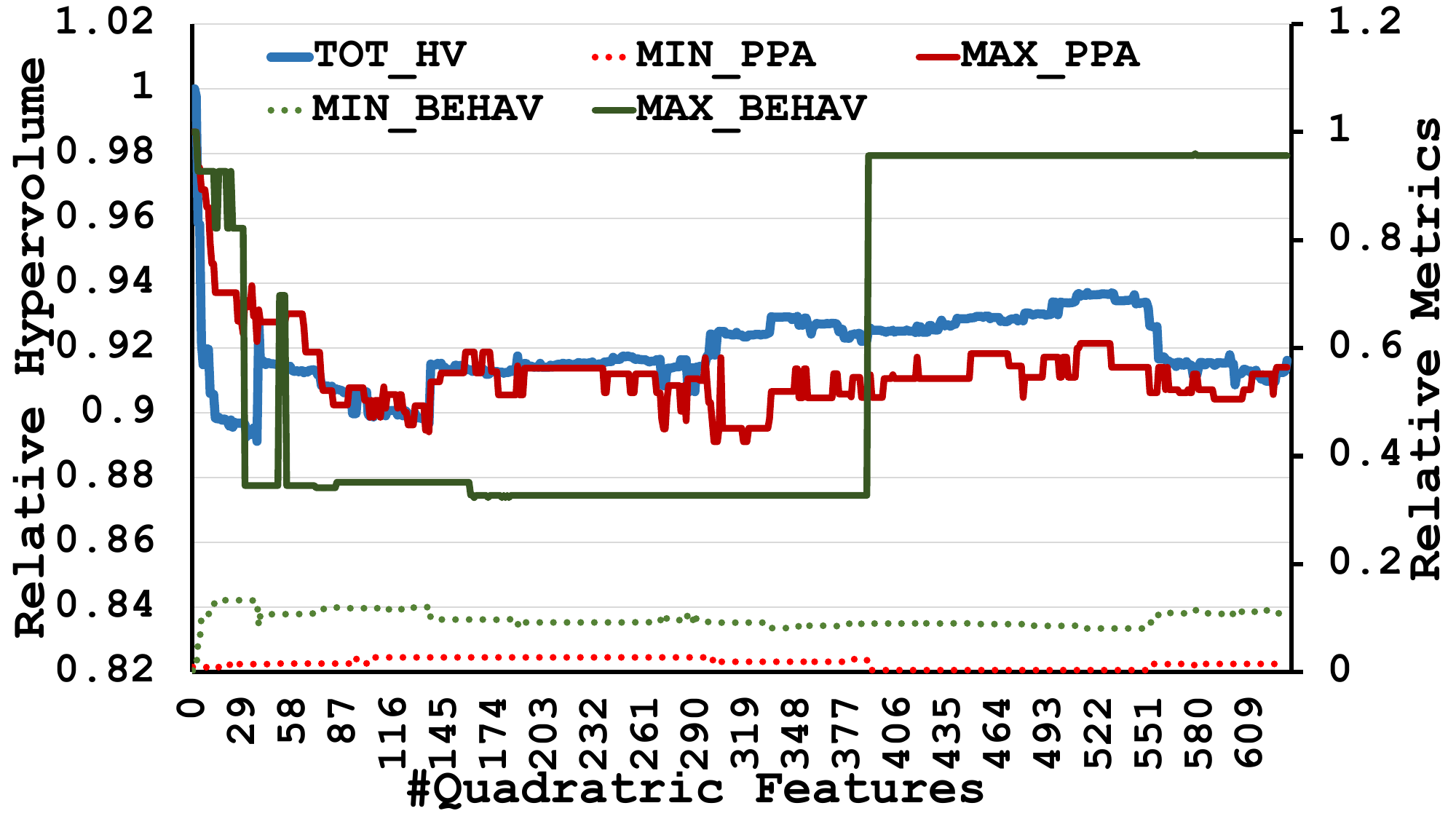}}
	\caption{
 PPF hypervolume (TOT\_HV) and the maximum and minimum PDPLUT(PPA) and AVG\_ABS\_REL\_ERR (BEHAV) reported in the solutions generated by the MaP problems using the polynomial regression models with an increasing number of quadratic terms
 }
	\label{fig:exp_map_ppf}
\end{figure}

\siva{
The design space exploration approach proposed in the current article comprises of generating a set of solutions to the constrained multi-objective optimization problem using \gls{map} and then using the solutions from the \gls{map}-based DSE to augment an evolutionary search. \autoref{fig:exp_map_models} shows the R\textsuperscript{2} score and \gls{mae} values (for TRAIN and TEST) for the \gls{pr} models used in formulating the \gls{map} problems for \gls{dse} of approximate signed $8 \times 8$ multipliers. 
The PPA and BEHAV correspond to PDPLUT and AVG\_ABS\_REL\_ERR respectively and each data point corresponds to adding one additional quadratic term. 
Hence, the graph plots results for problems with zero (MILP) to 630 quadratic terms, similar to the results for the signed $4 \times 4$ multipliers (shown in~\autoref{fig:motiv_2}), the R\textsuperscript{2} score for the training data increases with the addition of each quadratic terms.
}

\siva{
However, the TEST R\textsuperscript{2} score starts to decline after $\sim$200 terms. This can be attributed to the lack of the generalization of \gls{pr} models in predicting AVG\_ABS\_REL\_ERR and resulting in overfitting.
\autoref{fig:exp_map_ppf} shows the total hypervolume (TOT\_HV) of the \gls{ppf} and the minimum and maximum PDPLUT (MIN\_/MAX\_PPA) and AVG\_ABS\_REL\_ERR(MIN\_/MAX\_BEHAV) reported in the solutions of the \gls{map} problems corresponding to constraint scaling factor, $const\_sf$, of 0.5. 
As observed, the highest hypervolume is reported with the addition of the first few quadratic terms. Adding more terms results in increasing the complexity of the \gls{map} problems and may result in degraded results. 
However, using multiple problems allows us to generate a solution pool that can be used for further exploration.
}

\subsubsection{\gls{map}-augmented DSE}

\begin{figure}[t]
	\centering
    \subfloat[ $const\_sf$ is set to 0.2]{
    \scalebox{1.0}{
    \includegraphics[width=0.5 \columnwidth]{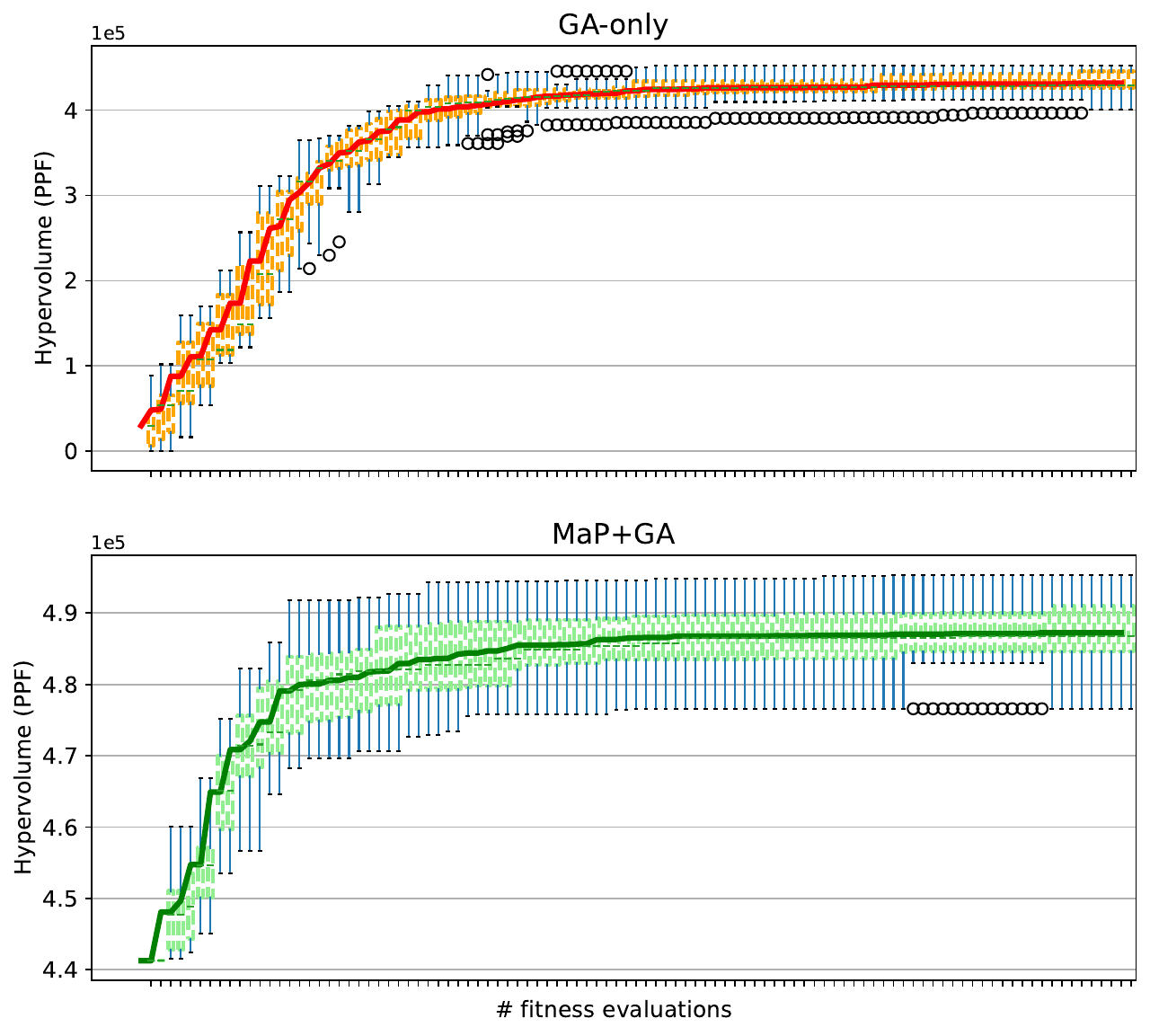}
    }}
    \subfloat[ $const\_sf$ is set to 0.5]{
    \scalebox{1.0}{
    \includegraphics[width=0.5 \columnwidth]{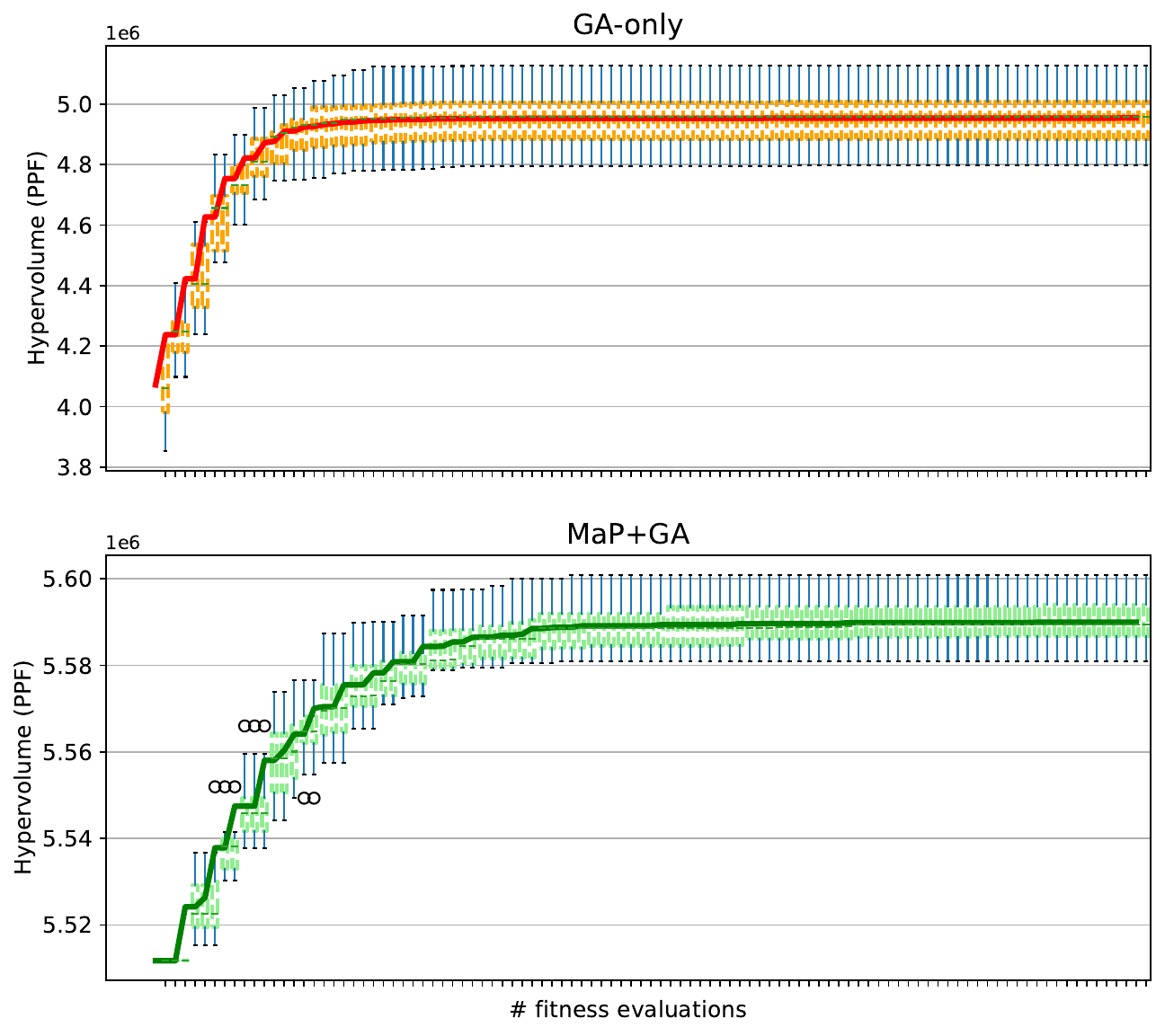}
    }}
    \\
    \subfloat[ $const\_sf$ is set to 0.8]{
    \scalebox{1.0}{
    \includegraphics[width=0.5 \columnwidth]{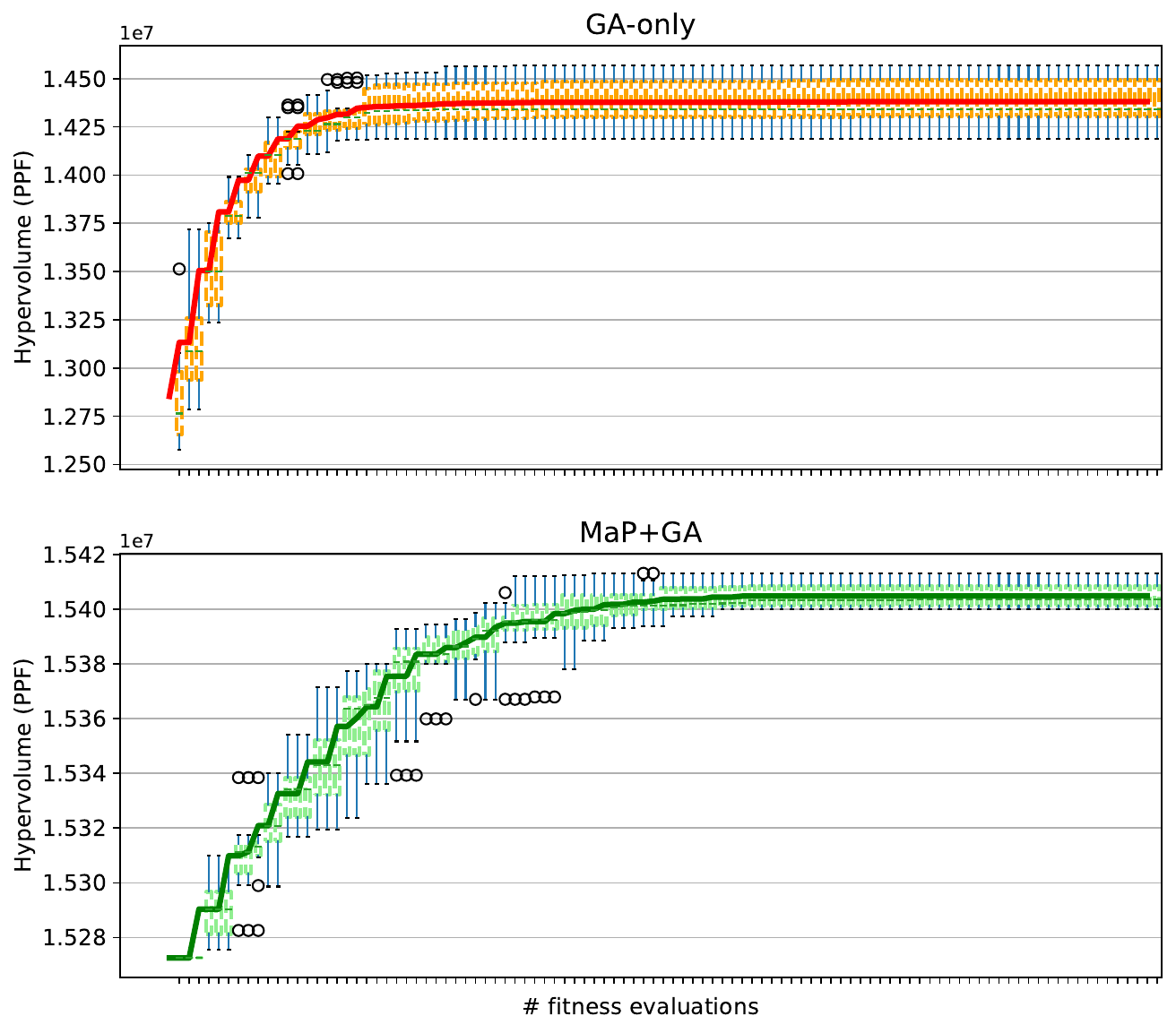}
    }}
    \subfloat[ $const\_sf$ is set to 1.0]{
    \scalebox{1.0}{
    \includegraphics[width=0.5 \columnwidth]{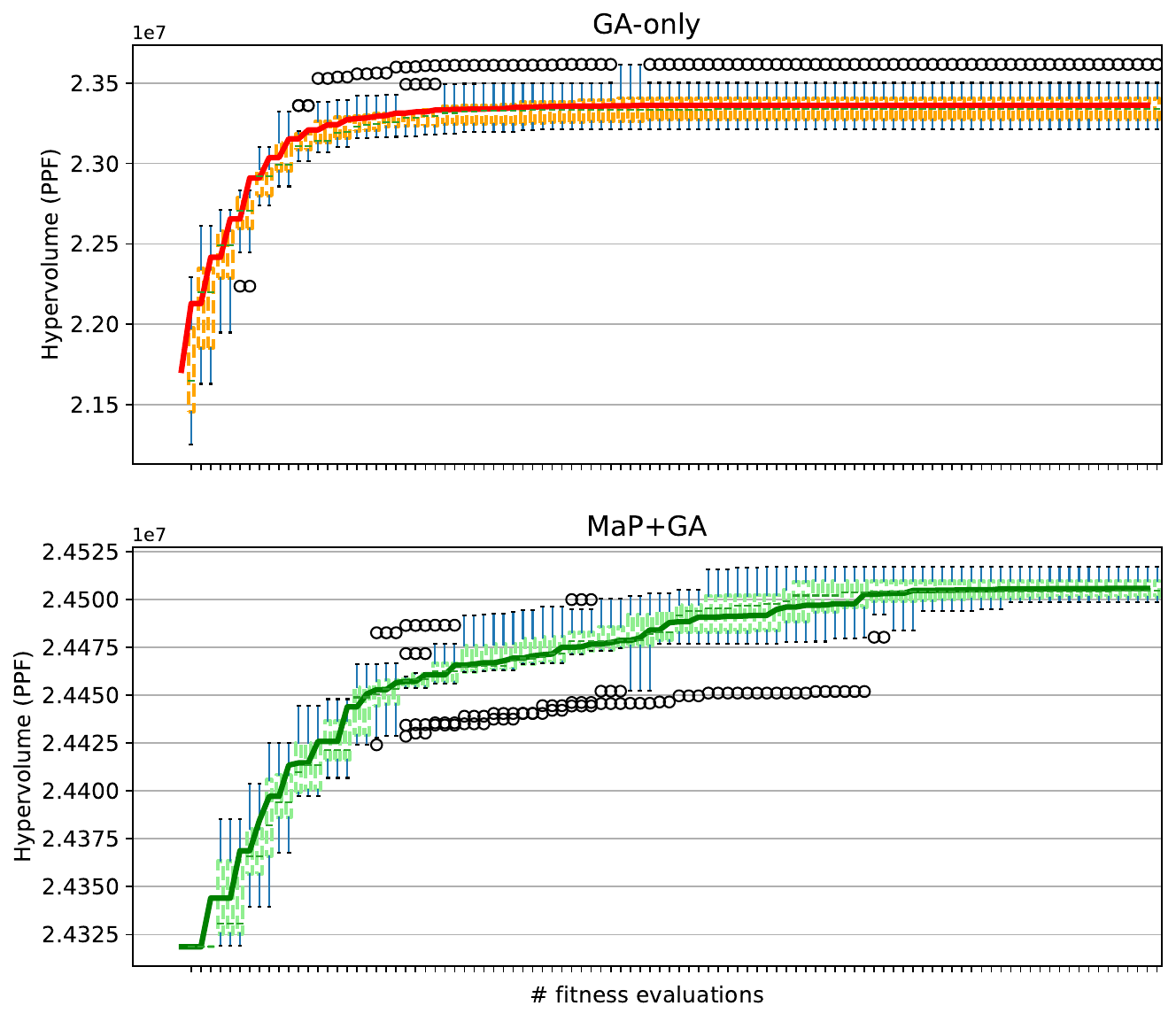}
    }}
    \\
    \subfloat[ $const\_sf$ is set to 1.2]{
    \scalebox{1.0}{
    \hspace*{-0.8cm}
    \includegraphics[width=1.1 \columnwidth]{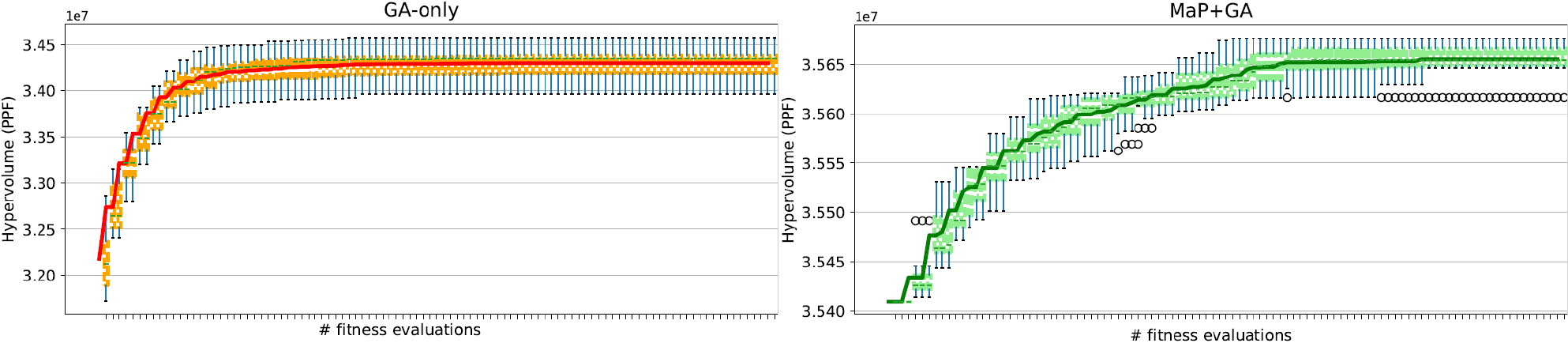}
    }}
	\caption{Progress of the distribution of the hypervolume across 10 different searches reported by generic GA (GA-only) and MaP-augmented GA (MaP+GA) search methods after an equivalent number of increasing fitness evaluations. The line traces the mean.}
	\label{fig:exp_prog_ga}
\end{figure}

\siva{
In addition to generating designs using \gls{map}, we also used a \gls{map}-augmented metaheuristic optimization. \autoref{fig:exp_prog_ga} shows the progression of the hypervolume in two methods of GA-based searches: (a) \textit{GA only}: problem agnostic evolution and (b) \textit{MaP+GA}: evolution with initial solutions generated using \gls{map}, in generating approximate signed $8 \times 8$ multipliers. 
Each of the sub-figures shows the progression for a different $const\_sf$. 
Each figure shows the distribution of the results for an increasing number of fitness evaluations across 10 runs with varying random seeding. As evident from the figure, the \gls{map}-augmented \gls{ga} shows much-improved hypervolume and continues improving till more iterations than the more generic GA-based search. It must be noted that the plots show the progression of the hypervolume of the \gls{ppf} and not the hypervolume of the \gls{vpf}. 
}

\begin{figure}[htb!]
	\centering
	\scalebox{1}{\includegraphics[width=0.9 \textwidth]{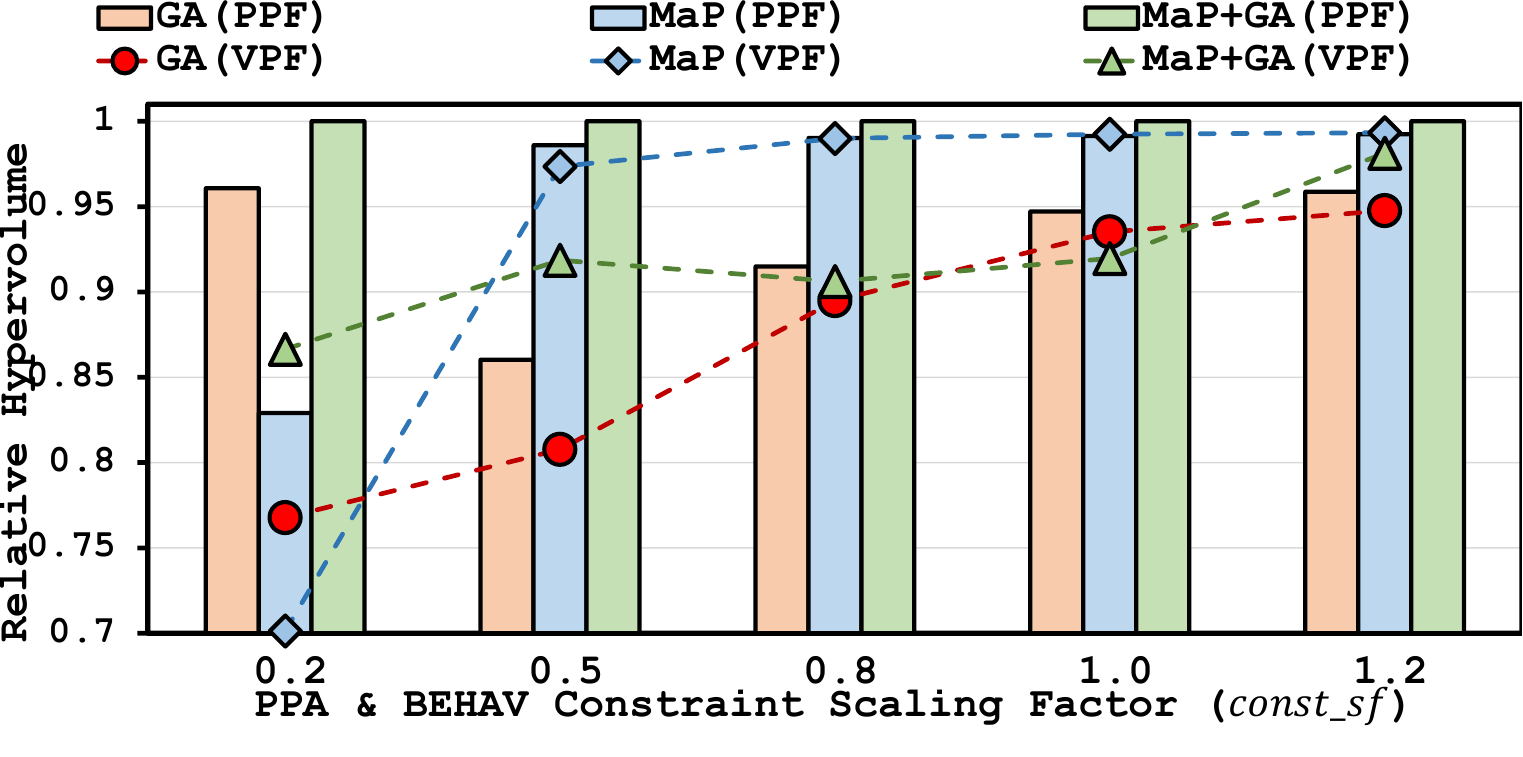}}
	\caption{Comparison of the relative hypervolume of the Pareto-front design points (PPF and VPF) obtained in the minimization of PDPLUT and AVG\_ABS\_REL\_ERR in approximate signed $8 \times 8$ multipliers for problems with different levels of maximum PPA and BEHAV constraints.}
	\label{fig:exp_overall_ga}
\end{figure}

\siva{
\autoref{fig:exp_overall_ga} shows the comparison of the final hypervolume obtained by three approaches---\gls{ga}, \gls{map}, \gls{map}+\gls{ga}---for different constraint scaling factors. While the bar plots in the figure show the \gls{ppf} hypervolume based on \gls{ml}-based estimation of PPA and BEHAV metrics of the Pareto-front configurations, the markers show the \gls{vpf}, based on metrics derived from the actual characterization of the Pareto-front configurations. The \gls{map}+\gls{ga} results in better \gls{ppf} hypervolume than GA-only across all scenarios. For loosely constrained problems, \gls{map}-based \gls{vpf} hypervolume is better/equivalent to other approaches. Overall, we observed up to 21\% improvements with \titleName-based methods compared to generic \gls{ga}.
}

\clearpage

\begin{figure}[htb!] 
    \centering
    \subfloat[ $const\_sf$ is set to 0.2]{
    \scalebox{1.0}{
    \includegraphics[width=0.5 \columnwidth]{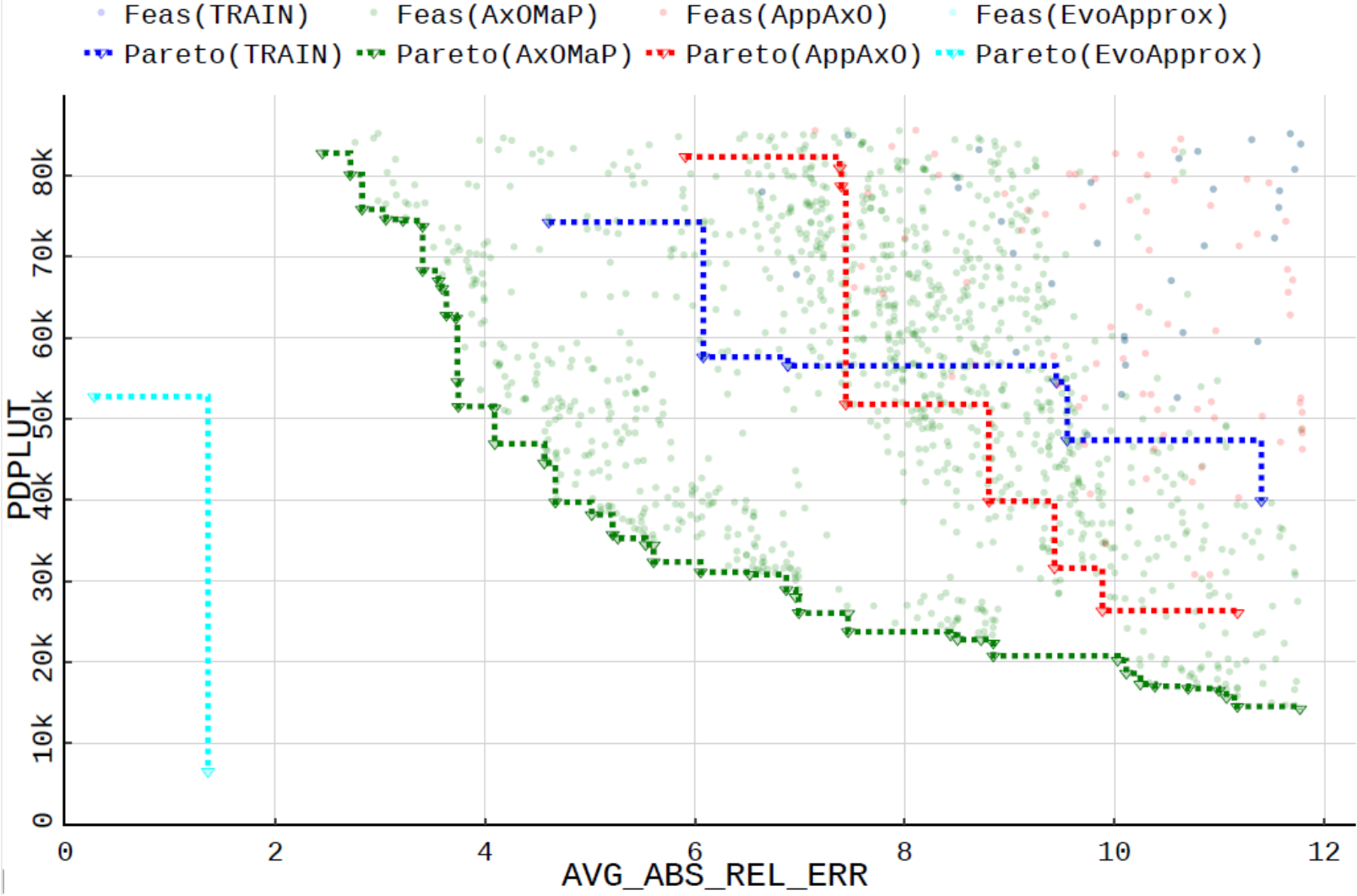}
    }}
    \subfloat[ $const\_sf$ is set to 0.5]{
    \scalebox{1.0}{
    \includegraphics[width=0.5 \columnwidth]{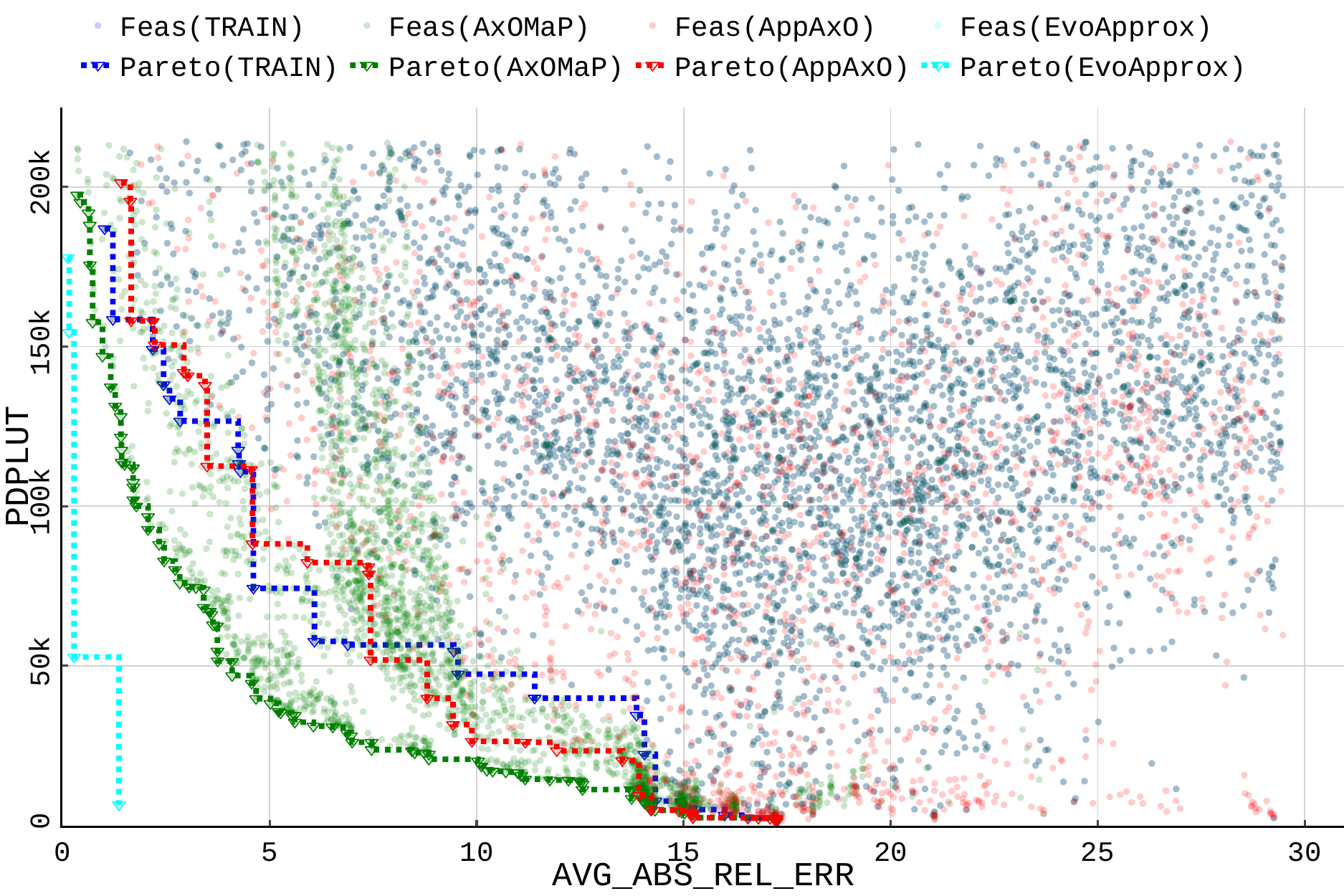}
    }}
    \\
\subfloat[ $const\_sf$ is set to 0.8]{
    \scalebox{1.0}{
    \includegraphics[width=0.5 \columnwidth]{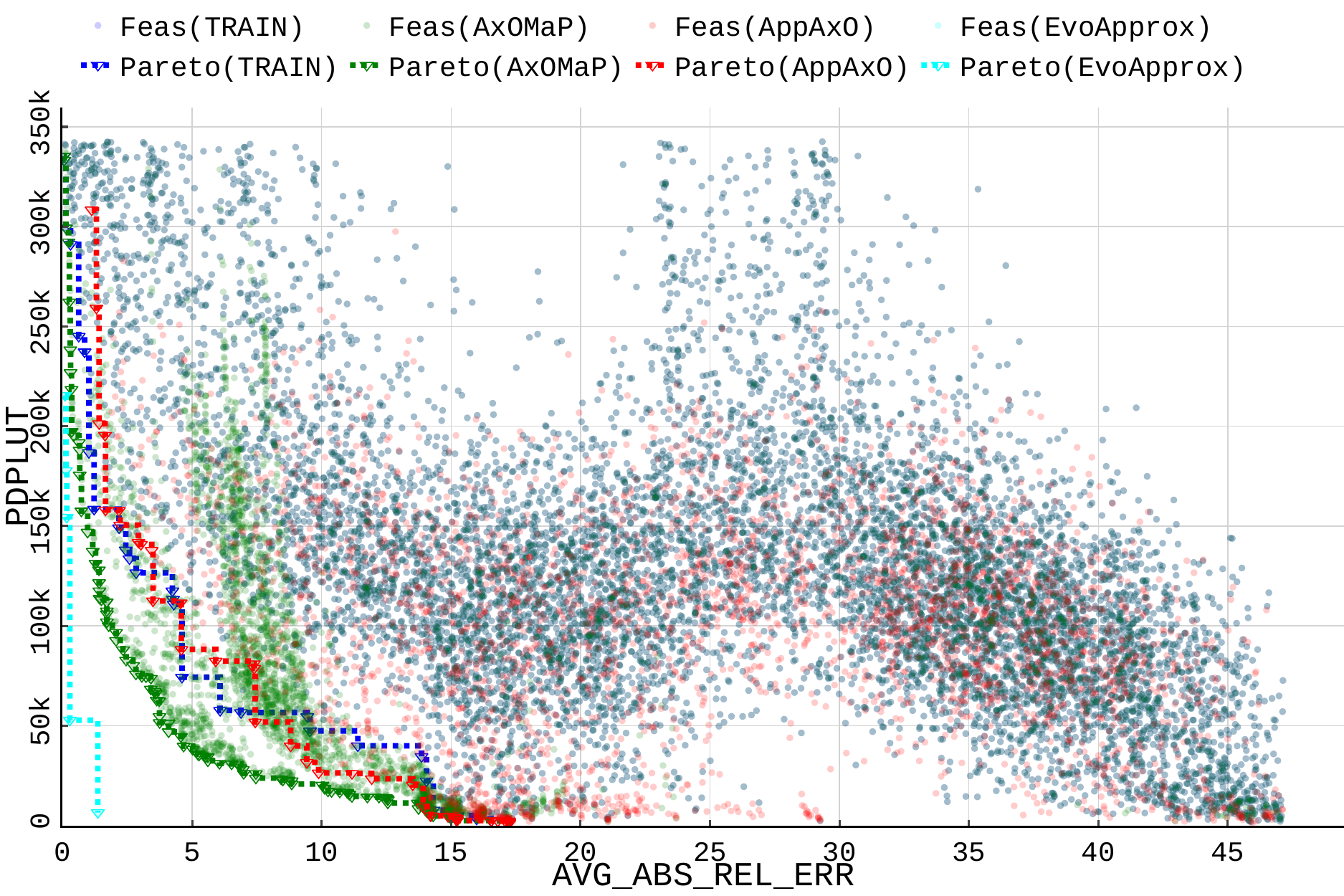}
    }}
    \subfloat[ $const\_sf$ is set to 1.0]{
    \scalebox{1.0}{
    \includegraphics[width=0.5 \columnwidth]{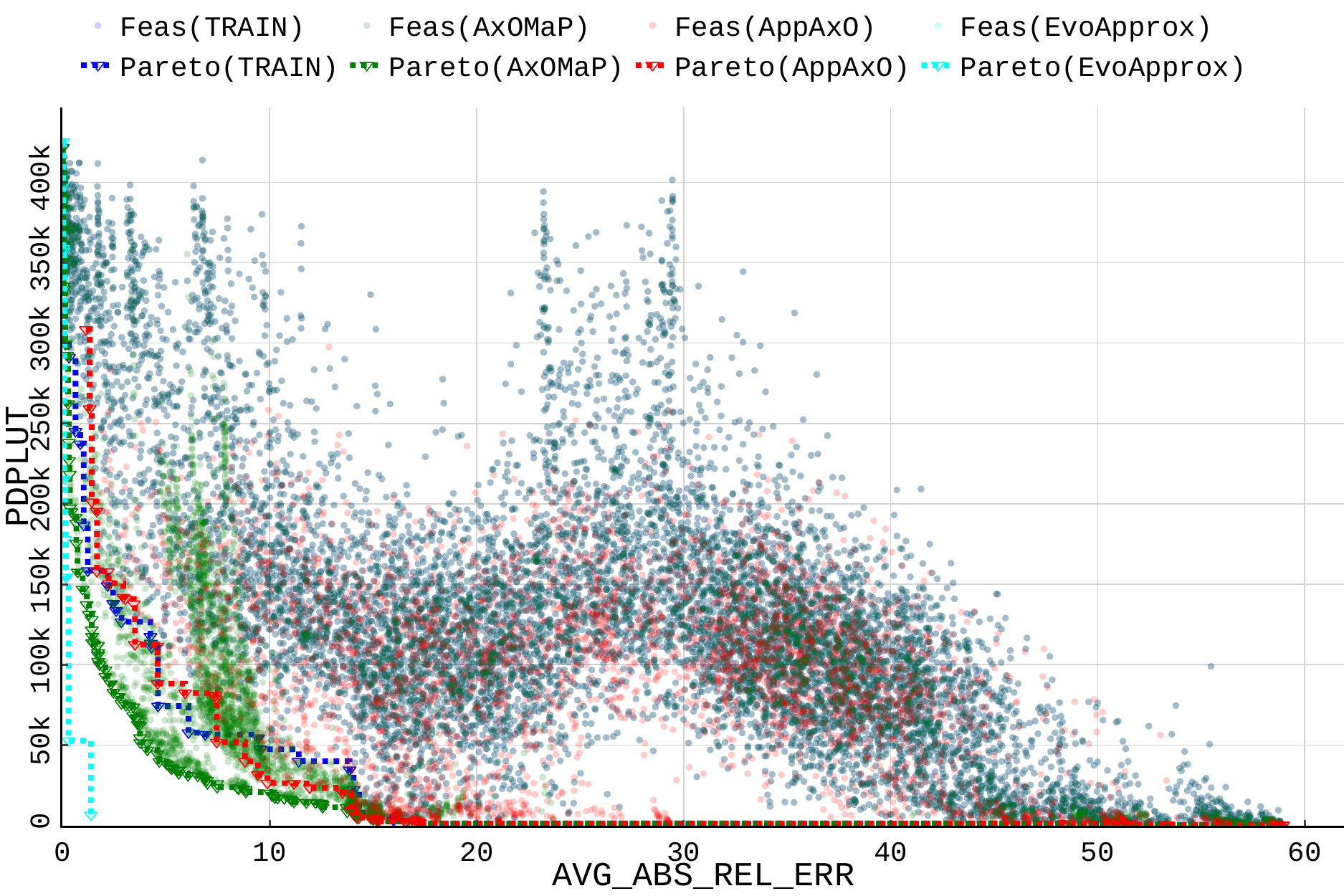}
    }}
    \\
    \subfloat[ $const\_sf$ is set to 1.2]{
    \scalebox{1.0}{
    \includegraphics[width=0.8 \columnwidth]{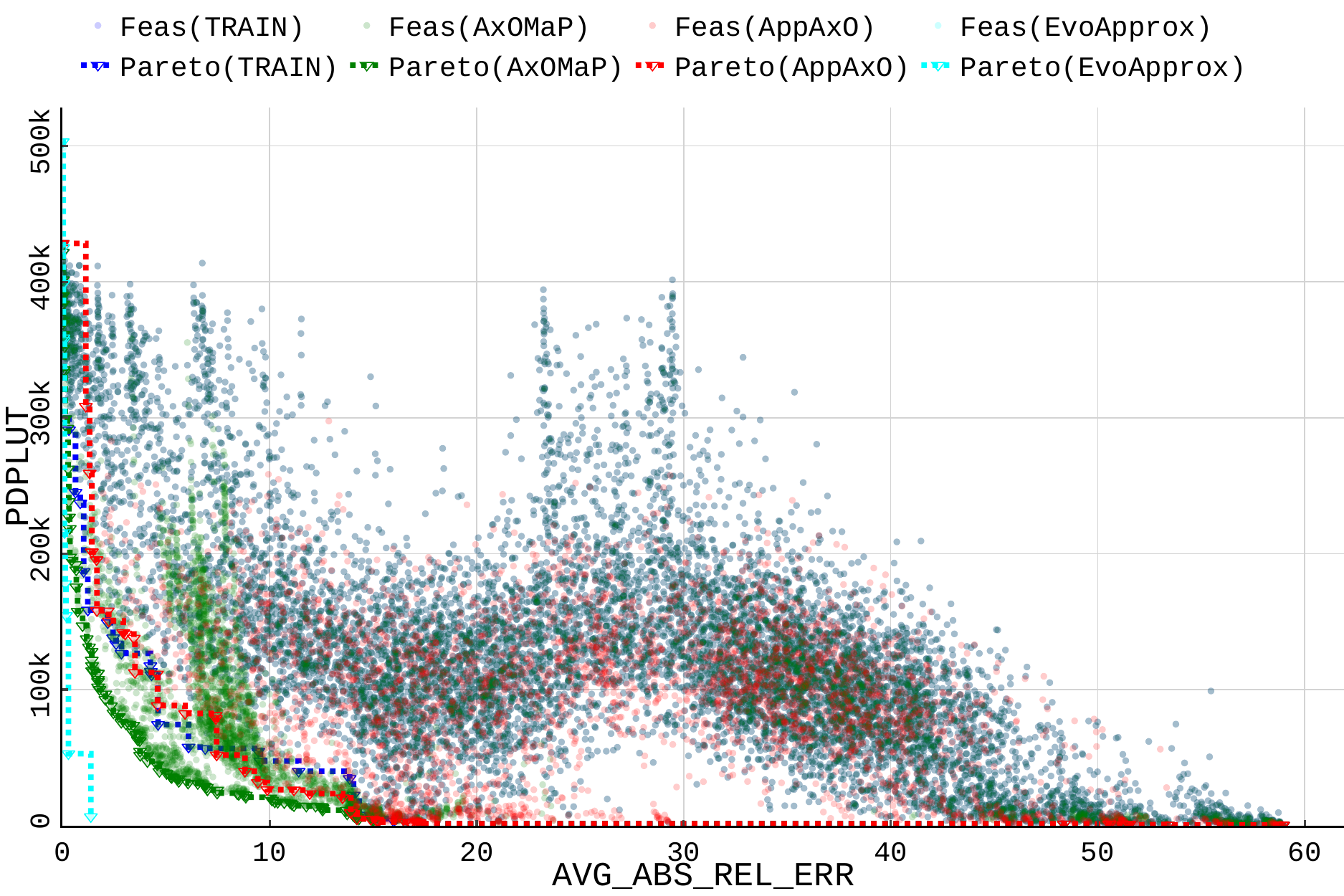}
    }}
\caption{Comparison of the Pareto-front obtained by \titleName~ to that reported in \textit{AppAxO}~\cite{ullah2022appaxo} and \textit{EvoApprox}~\cite{evoapprox16} for approximate signed $8 \times 8$ multipliers.}
  \label{fig:soa_pareto_op} 
\end{figure}

\subsection{Comparison with State-of-the-art}
\subsubsection{Operator-level DSE}

\begin{figure}[t]
	\centering
	\scalebox{1}{\includegraphics[width=0.9 \columnwidth]{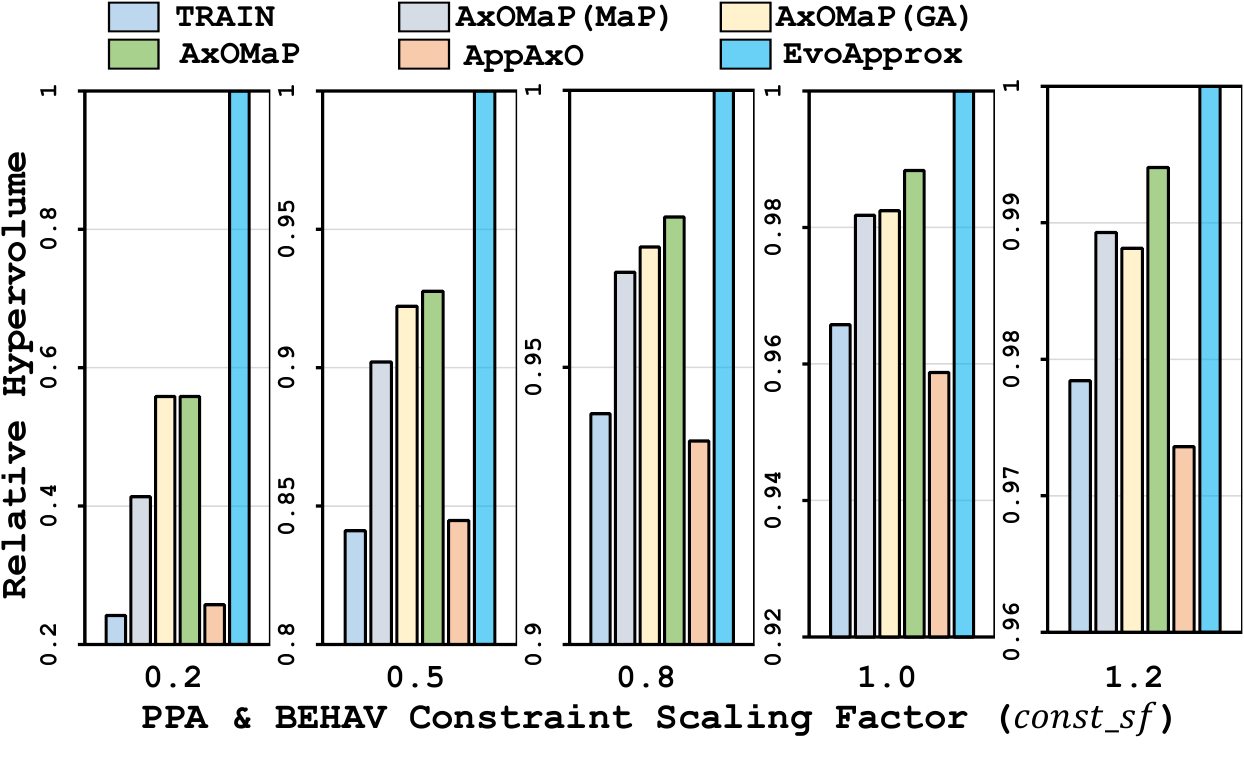}}
	\caption{Comparison of the Pareto-front hypervolume obtained with \titleName~compared to that in related works, for different DSE problems obtained by varying the $const\_sf$ values}
	\label{fig:soa_hyp_op}
\end{figure}

\siva{
In order to compare with related state-of-the-art approaches to designing FPGA-based \glspl{axo}, we evaluated the proposed methods for both operator-level DSE and application-specific DSE for synthesizing novel approximate signed 8-bit multipliers.
\autoref{fig:soa_pareto_op} shows the different design points and the Pareto-fronts obtained using the training data in \titleName, \gls{dse} results of \titleName~and the reported results of \textit{EvoApprox}~\cite{evoapprox16} and \textit{AppAxO}~\cite{ullah2022appaxo}. 
Each sub-figure shows the Pareto-front for a different value of $const\_sf$.
While in \textit{EvoApprox}, \gls{asic}-optimized logic is synthesized and implemented on \glspl{fpga}, the other methods involve optimizing for \glspl{fpga}. With \titleName~ we report considerably better Pareto-front than \textit{AppAxO}. 
}

\siva{
The resulting hypervolume of the solutions at different constraint scaling is shown in \autoref{fig:soa_hyp_op}. All the results correspond to the \gls{vpf}. As can be seen in the figure, \titleName~ methods, both \gls{map} and \gls{map}+\gls{ga}, result in a better quality of results than those reported in \textit{AppAxO}. We report 116\%, 9.8\%, 4.3\%, 3.1\% and 2.1\% higher hypervolume than \textit{AppAxO} for $const\_sf$ values of 0.2, 0.5, 0.8, 1.0, and 1.2 respectively.
As evident from both \autoref{fig:soa_pareto_op} and \autoref{fig:soa_hyp_op}, the largest improvement over the designs reported in \textit{AppAxO} occurs in the case of more tightly constrained search. The approximate design configurations obtained from the \gls{map}-based \gls{dse}, being solutions of constrained problems themselves, aid in generating better solutions under such tighter constraints. As seen in \autoref{fig:soa_pareto_op}, with rising tolerance to error (for less tightly constrained problems) more designs are generated with higher cost to accuracy while resulting in lower PDPLUT. It can be noted from \autoref{fig:soa_hyp_op}, that we report considerable improvements over \textit{AppAxO}~\cite{ullah2022appaxo}, the other methodology implementing FPGA-specific LUT-level optimizations.
}

\clearpage
\subsubsection{Application-specific DSE}
\begin{figure}[t]
	\centering
	\scalebox{1}{\includegraphics[width=0.8 \textwidth]{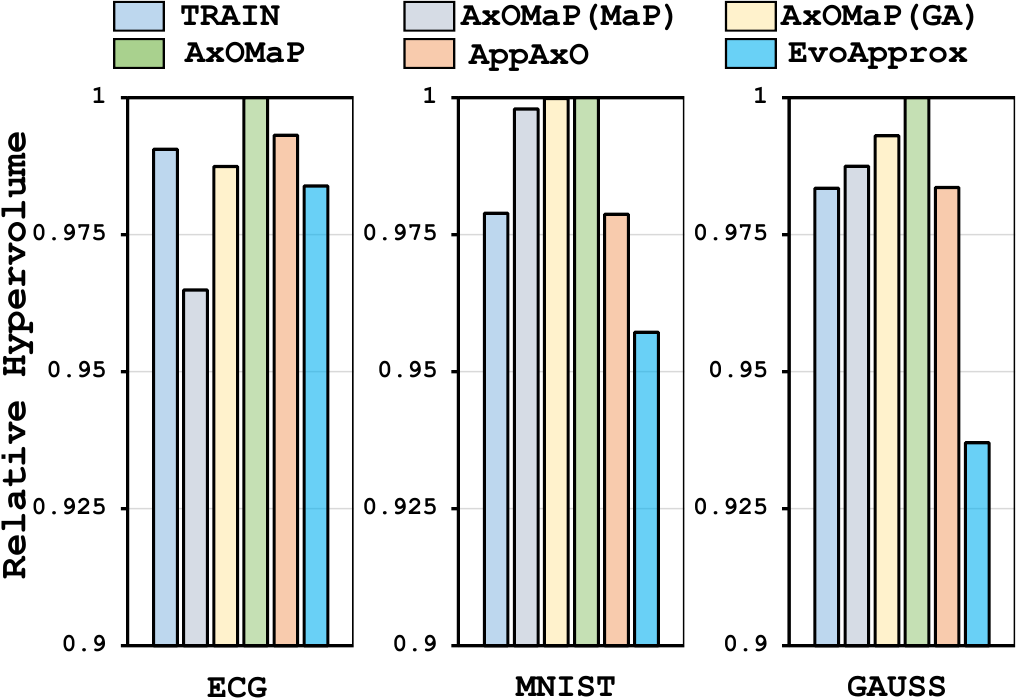}}
     \caption{Pareto-front hypervolume obtained for the application-specific search of approximate signed $8\times 8$ multipliers.}
	\label{fig:soa_hyp_app}
\end{figure}
\siva{
As discussed in \textit{AppAxO}, the operator model used does not allow the exploration of the \textit{init values} in the LUTs, as is possible in \textit{EvoApprox} designs. Therefore, the results at the operator-level optimization are worse than that reported with \textit{EvoApprox}. 
However, the ability to generate more novel designs, by removing a different subset of LUTs, allows the LUT-based optimization methodologies to adapt to any inherent tolerance of different applications. Using the designs generated by \textit{EvoApprox}, on the other hand, limits the exploration to a fixed set of approximate designs.
}

\begin{figure}[htb!] 
    \centering
    \subfloat[ $const\_sf$ is set to 0.5]{
    \scalebox{1.0}{
    \includegraphics[width=0.45 \columnwidth]{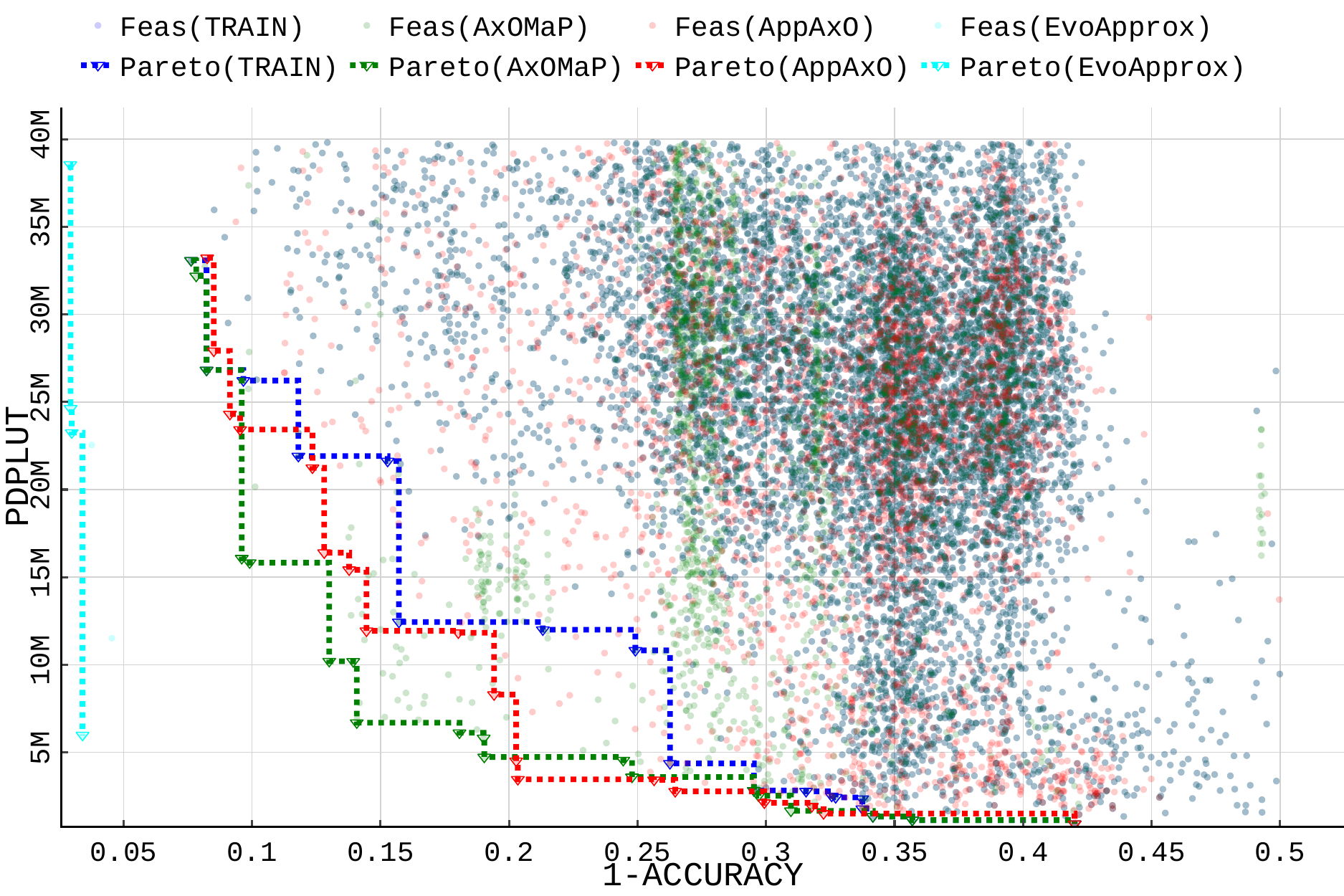}
    }}
    \subfloat[ $const\_sf$ is set to 0.8]{
    \scalebox{1.0}{
    \includegraphics[width=0.45 \columnwidth]{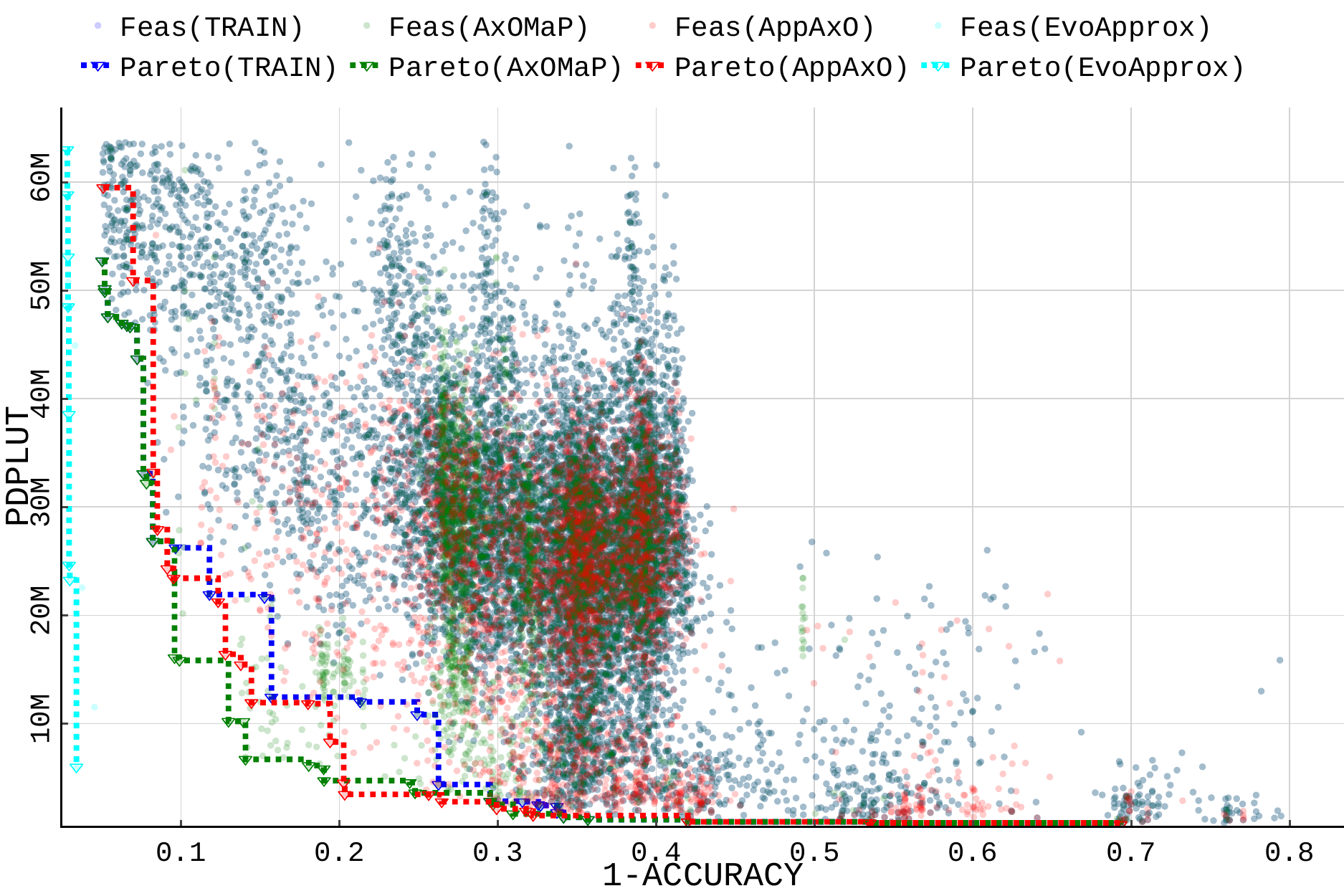}
    }}
\caption{Pareto-fronts for using approximate signed $8 \times 8$ multipliers in the LPF of ECG peak detection.}
  \label{fig:soa_pareto_ecg} 
\end{figure}

\begin{figure}[htb!] 
    \centering
    \subfloat[ $const\_sf$ is set to 0.5]{
    \scalebox{1.0}{
    \includegraphics[width=0.45 \columnwidth]{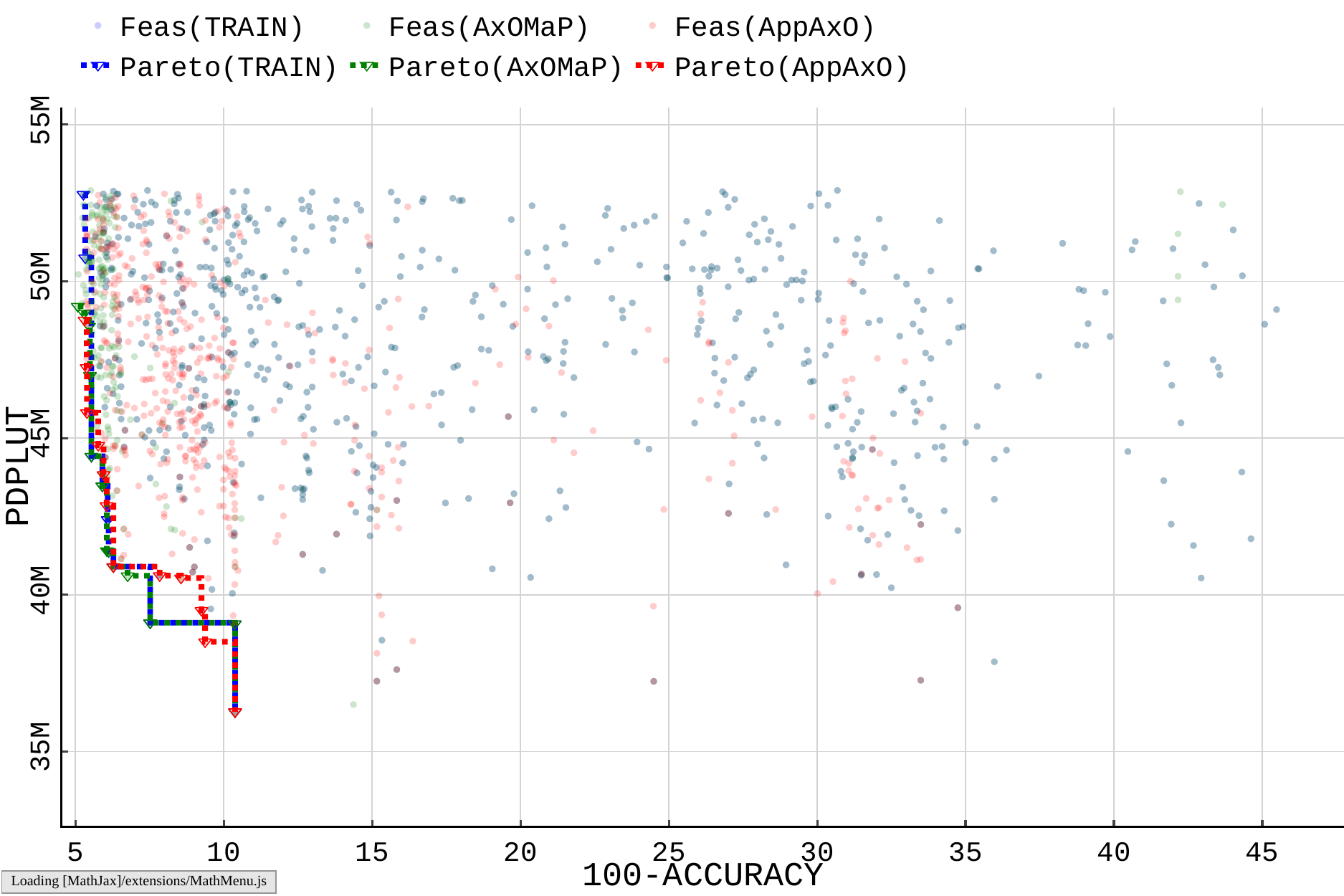}
    }}
    \subfloat[ $const\_sf$ is set to 0.8]{
    \scalebox{1.0}{
    \includegraphics[width=0.45 \columnwidth]{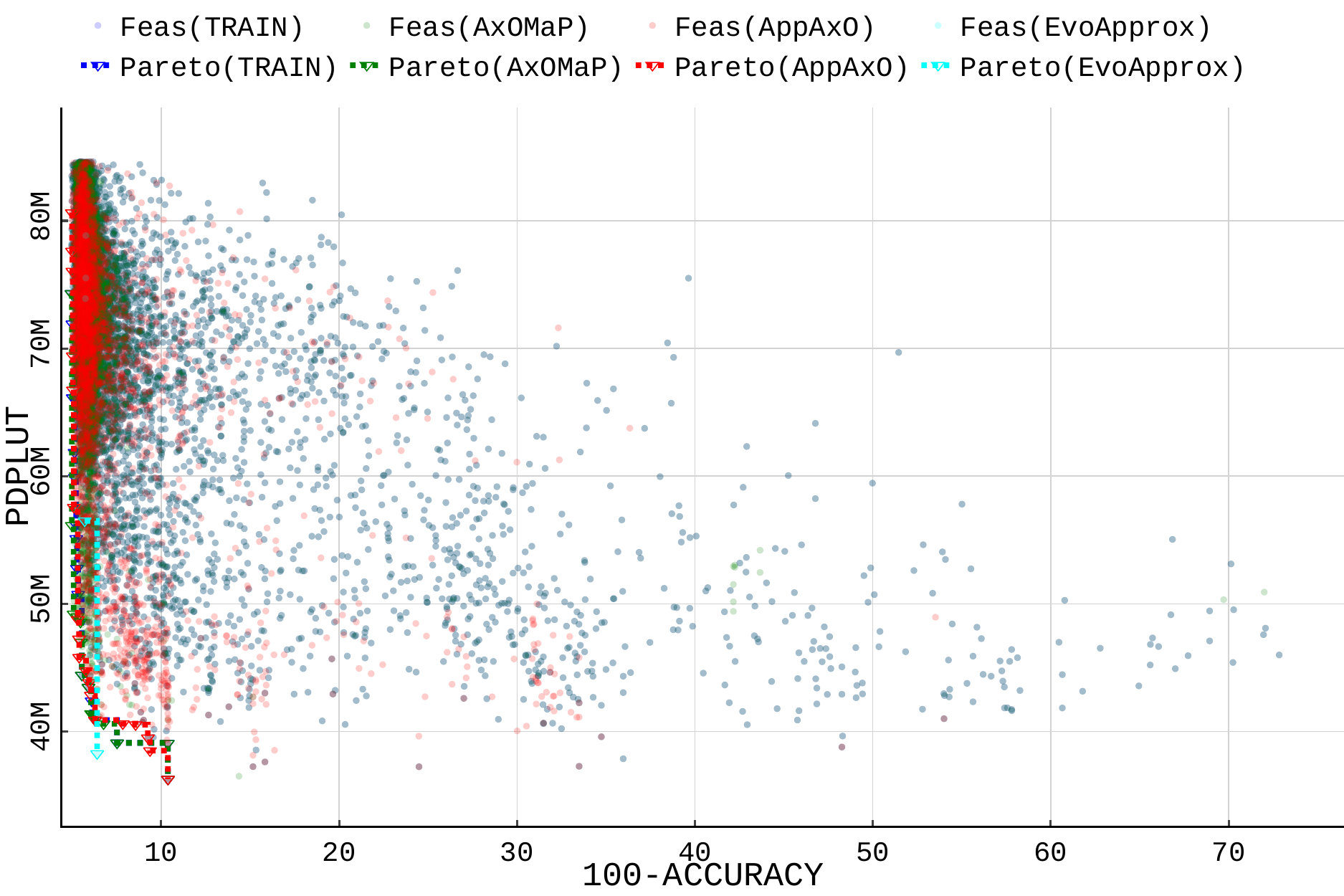}
    }}
\caption{Pareto-fronts for using approximate signed $8 \times 8$ multipliers in the MNIST digit classifier.}
  \label{fig:soa_pareto_mnist} 
\end{figure}

\begin{figure}[htb!] 
    \centering
    \subfloat[ $const\_sf$ is set to 0.5]{
    \scalebox{1.0}{
    \includegraphics[width=0.45 \columnwidth]{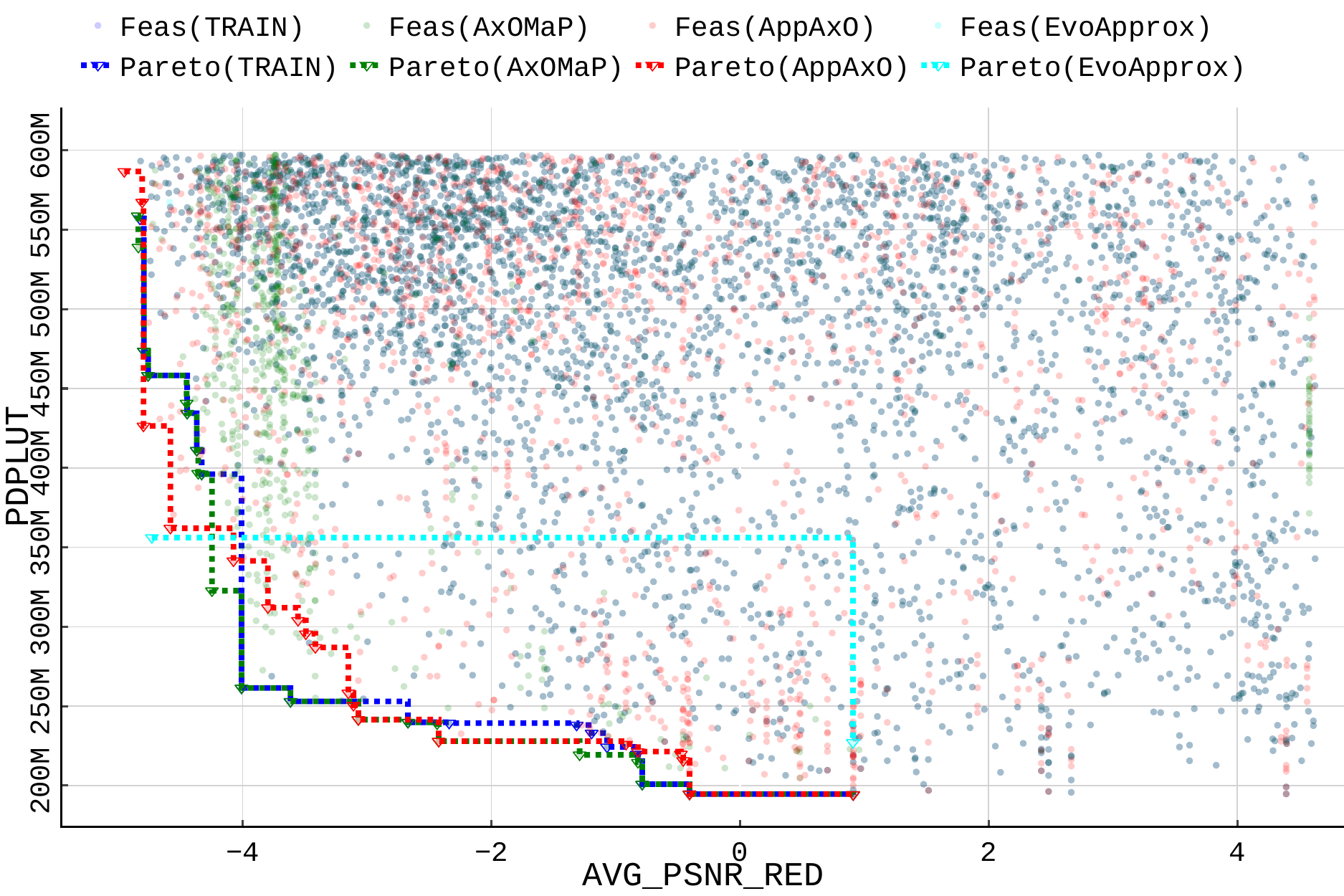}
    }}
    \subfloat[ $const\_sf$ is set to 1.2]{
    \scalebox{1.0}{
    \includegraphics[width=0.45 \columnwidth]{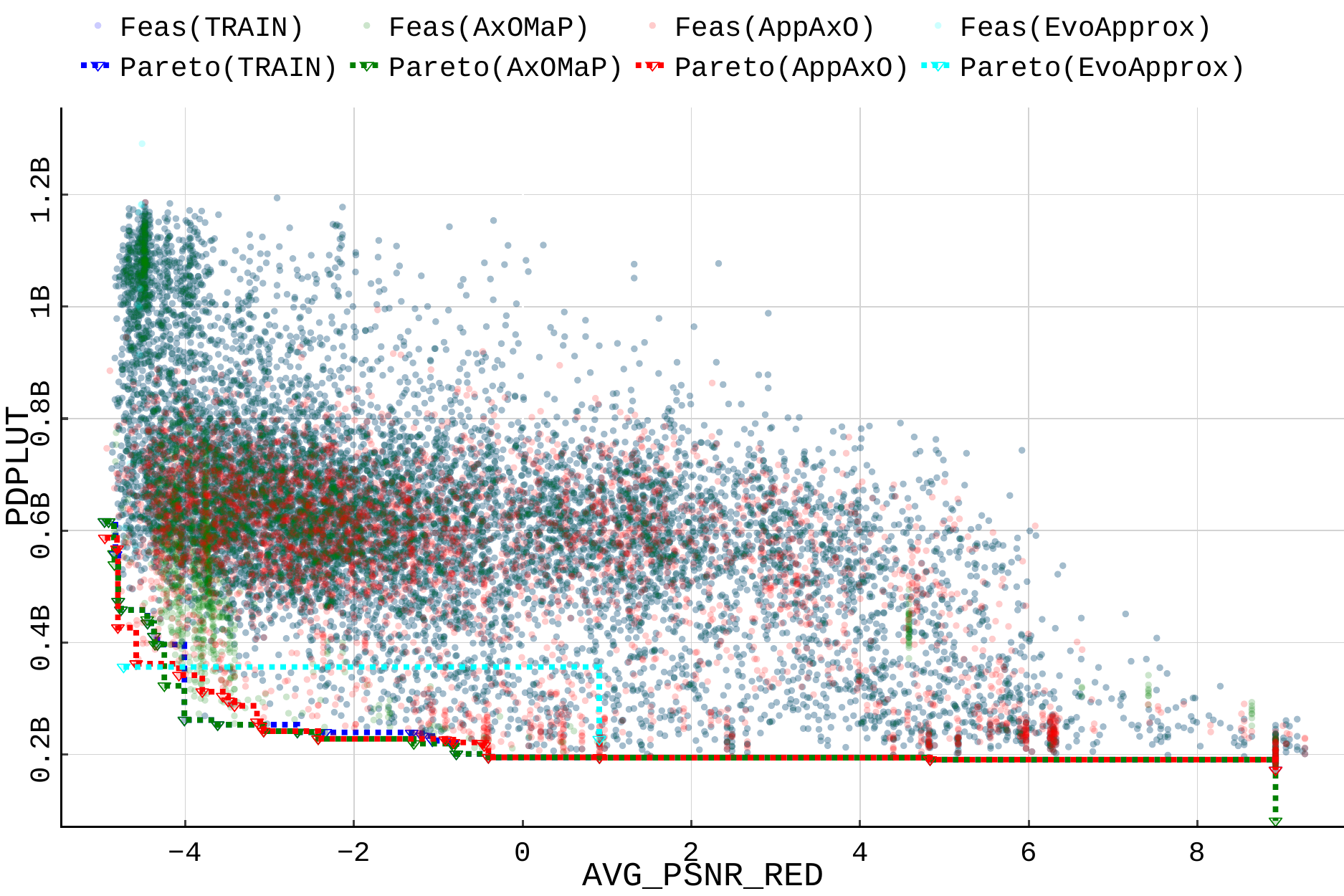}
    }}
\caption{Pareto-fronts for using approximate signed $8 \times 8$ multipliers in 2D  gaussian smoothing filter.}
  \label{fig:soa_pareto_gauss} 
\end{figure}

\siva{
Consequently, for application-specific \gls{dse} we report much better quality of results than both \textit{EvoApprox} and \textit{AppAxO}. \autoref{fig:soa_hyp_app} shows the comparison of the resulting hypervolume in the application-specific \gls{dse} for three different applications listed in~\autoref{table:exp_app}.
While the figure shows the results for an almost unconstrained search (with $const\_sf$ set to 1.5), with other values of $const\_sf$ we observe up to 21\%, 13\%, and 27\% better hypervolume than related works for ECG, MNIST, and GAUSS respectively. 
\autoref{fig:soa_pareto_ecg}, \autoref{fig:soa_pareto_mnist} and \autoref{fig:soa_pareto_gauss} show the Pareto front of design points obtained for each of the application for two values of $const\_sf$.
In the case of ECG, the benefits with \textit{AppAxO} and \titleName~ derive from generating designs with large costs to accuracy. For MNIST, \titleName~ generates designs with limited improvements over the \titleName(TRAIN) and \textit{AppAxO}. It can be noted that none of the designs of \textit{EvoApprox} result in feasible designs at tighter constraints ($const\_sf=0.5$). Similarly, for GAUSS only two designs from \textit{EvoApprox} are feasible, with only one design providing any useful benefit (AVG\_PSNR\_RED < 0).
}
\clearpage

\clearpage
\section{Conclusion}
\label{sec:conc}


\siva{
Approximate computing is one of the more widely researched design approaches for enabling future resource-constrained smart systems. The inherent noise-tolerant behavior of emerging applications such as ML, data mining, etc. provides a conducive environment to approximate computing. However, extracting the maximum benefits with approximate systems for such applications requires \gls{dse} approaches that enable a high level of application-specific search, at various
levels of the system stack.
To this end, the current article, we present a novel methodology for synthesizing FPGA-based approximate operators. The proposed approach leverages the statistical analysis of characterization data to improve the efficacy of prevalent DSE methods. Using the proposed methods, we report up to 116\% and 27\% improved designs than state-of-the-art approaches for operator-level and application-specific DSE respectively. The proposed methodology can be extended to use more sophisticated analysis methods and operator models for improving the DSE.
}     

\bibliographystyle{ACM-Reference-Format}
\bibliography{full_ref}

\end{document}